\documentclass[preprint]{aastex61}

\usepackage{natbib}
\usepackage[section]{placeins}
\usepackage{color}
\usepackage{amsmath}

\newcommand{\lau}[1]{{{#1}}}

\shorttitle{Interacting $\Omega$-loops}
\shortauthors{Jouve, Brun \& Aulanier}

\begin{document}

\bibliographystyle{plainnat}

\title{Interactions of twisted $\Omega$-loops in a model solar convection zone}

\author{L. Jouve}
\affil{IRAP, Universit\'e de Toulouse, CNRS, UPS, CNES, Toulouse, France, 14 Avenue Edouard Belin, 31400 Toulouse, France}

\email{ljouve@irap.omp.eu}

\author{A. S. Brun}
\affil{Laboratoire AIM, CEA/DRF-CNRS/INSU-Universit\'e 
Paris Diderot, IRFU/DAp, 91191 Gif sur Yvette, France}

\author{G. Aulanier}
\affil{LESIA, Observatoire de Paris, CNRS, UPMC, Universit\'e Paris-Diderot, 5 Place Jules Janssen, 92190 Meudon Cedex, France}

\begin{abstract}
This study aims at investigating the ability of strong interactions between magnetic field concentrations during their rise through the convection zone to produce complex active regions at the solar surface. To do so, we perform numerical simulations of buoyant magnetic structures evolving and interacting in a model solar convection zone. We first produce a 3D model of rotating convection and then introduce idealized magnetic structures close to the bottom of the computational domain. These structures possess a certain degree of field line twist and they are made buoyant on a particular extension in longitude. The resulting twisted $\Omega$-loops will thus evolve inside a spherical convective shell possessing large-scale mean flows. We present results on the interaction between two such loops with various initial parameters (mainly buoyancy and twist) and on the complexity of the emerging magnetic field. In agreement with analytical predictions, we find that if the loops are introduced with opposite handedness and same axial field direction or same handedness but opposite axial field, they bounce against each other. The emerging region is then constituted of two separated bipolar structures. On the contrary, if the loops are introduced with the same direction of axial and peripheral magnetic fields and if sufficiently close, they merge while rising. This more interesting case produces complex magnetic structures, with a high degree of non-neutralized currents, especially when the convective motions act significantly on the magnetic field. This indicates that those interactions could be good candidates to produce eruptive events like flares or CMEs.
\end{abstract}


\section{Introduction}

The Sun displays a large spectrum of magnetic phenomena on its surface and extended corona.
From sunspot emergence, to the formation of active regions and prominences and to the eruptions, flares and coronal mass ejections (CME), all these magnetic structures play a key role
in determining the overall temporal and spatial variability of our Sun. The organisation and geometry of magnetic fields play a key role
to power these events. These fields owe their origin to the \lau{dynamo mechanism} operating deep inside the solar convective layer.
Of particular importance to characterise the Sun's magnetic activity, is the link between dynamo and flux emergence and how complex topological structures 
are created and emerge to then lead to the \lau{intricate} configuration of the solar surface magnetic fields.
Statistical studies such as in \cite{2000ApJ...540..583S} have shown that there is a clear correlation between
the occurrence of intense flares and the complex magnetic topology of emerged sunspots. They showed that so-called $\delta$-spots with
mixed positive and negative polarities and circumvoluted polarity inversion lines (PIL)  are more prone
to eruption than simpler, regular and single dominated polarity $\alpha$-spot. 
There has thus been over the last decades a keen interest to understand through what internal processes do complex sunspots emerge at the solar surface. 
For instance \cite{2014SoPh..289.3351T} have studied, both observationally and through local 3-D MHD 
numerical simulations, the active region AR11558 that exhibited a complex magnetic topology and produced many intense flares. 
They compared the surface magnetic signatures of either one axially modulated flux rope or the interaction of two flux ropes 
to see if one scenario would explain the complex surface observation better. They concluded that convectively modulated flux 
ropes seemed to be the favoured scenario for that particular active region. Likewise, \cite{2010SoPh..261..127C} have studied in detail region AR10501 
in order to make a connection between its helicity and that inverted from 
the subsequent geoeffective magnetic cloud (MC) observed on 20 November 2003. They showed that even though the global magnetic helicity of the AR
was of opposite sign of that of the magnetic clouds, a small localised region within it, did possess the correct sign and was thus likely the location of the erupting event.

These studies, among many others, rest on 3-D MHD numerical simulations of flux rope emergence or surface coronal dynamics.
Nowadays, most of these studies also take into account the influence of turbulent convective motions. Indeed the solar surface is
paved with convection patterns that further help making the emerging flux complex and time evolving,
by buffeting magnetic field into the down flow lanes surrounding the up flows.
There are thus been many numerical experiments that have sought to model flux emergence in either
adiabatic or convective layers. \lau{Originally these studies have used the thin flux tube approximation \citep{1981A&A....98..155S, 1982A&A...106...58S}, studying the rise trajectory of tubes and tilt angle of emerging regions as a function of field strength, rotation rate and initial thermodynamic equilibrium \citep{1987ApJ...316..788C,1993A&A...272..621D}. Two dimensional MHD simulations were then performed in Cartesian or spherical geometries.
One of the main results of these studies was that a certain amount of twist was needed for the flux rope
to emerge as a single entity otherwise it would split because of the generation of vorticity lobes around the apex of the tube \citep{1998ApJ...492..804E}.}
The effect of rotation was also shown to result in an asymmetric rise, with the leading leg being more vertically oriented
than the trailing one (see \cite{1995ApJ...441..886C} and recent reviews by \citet{2009LRSP....6....4F} and \citet{2014LRSP...11....3C}).
Over the last two decades full 3-D MHD numerical simulations have now become the norm. There too, local
Cartesian simulations \citep{1999A&A...352L..79D,2003ApJ...582.1206F,2004ApJ...612..557A,2007A&A...467..703C,2008ApJ...679..871M,2010ApJ...720..233C,2014ApJ...785...90R,2015ApJ...814....2M} and global spherical simulations that can account for large scale flows and the latitudinal dependence of the Coriolis force \citep{2007AN....328.1104J,2008ApJ...676..680F,2009ApJ...701.1300J,Jouve3,2013ApJ...772...55P} have been performed. 
Most have focused on the emergence of a single isolated axisymmetric flux rope (including kink-unstable ones) and how convection and a realistic atmosphere influence their emerging properties. \lau{Some have been using idealised setups \citep{2006A&A...460..909M,2009A&A...503..999H,2011PASJ...63..407T,2012ApJ...745...37F,2013ApJ...778...42A}, while others have modeled the 
solar surface layers in detail considering for instance radiative transfer and the presence of an overlying corona \citep{2007A&A...467..703C,2008ApJ...679..871M,2010ApJ...720..233C,2015ApJ...813..112T,2015ApJ...814....2M}. }In the most ``solar-like" settings the formation of active regions has been studied in detail (including the formation of a penumbra or of the Evershed effect) and how they impact the corona above \lau{(see review on sunspot simulations by \citet{Rempel11})}.

\lau{In most calculations of a flux tube rising through the solar convection zone or into the atmosphere, the embedded flux rope can either be buoyant all along its main axis or chosen to be buoyant only in a limited region, leading
to the formation of $\Omega$-loops.} In the latter case, some gas can drain towards the anchored
area, thus modifying the overall dynamics. One important consequence is that $\Omega$-loops should require less twist to emerge
coherently \citep{2004ApJ...612..557A}. It is worth noting that in \cite{2014SoPh..289..441N} more than 130 self-consistently generated 
emerging $\Omega$-loops have been found in a dynamo simulation \citep[see also][]{2014ApJ...789...35F}, setting the stage for linking magnetic flux emergence and cyclic dynamo action in a so-called {\it spot-dynamo} framework \citep{2015SSRv..196..101B}.
However only few studies have looked at the interaction of multiple loops and how
they can or not explain the most complex sunspot groups or active regions.
There is for instance the work of \cite{Linton} 
in an idealised Cartesian setup.They showed that among the four possible configurations
for the combination of twist and orientation of the axial field in the flux ropes only some of them yield complex
interactions. This depends mostly on the ability of the two flux ropes to reconnect or not as they evolve. These different interactions were also studied in the same kind of setup by \cite{Sakai92} and in a more realistic situation mimicking reconnecting coronal loops by \cite{Ozaki97}.
In \cite{1998ApJ...493..480F,2007A&A...466..367A}, the interaction of two flux ropes in
a Cartesian stratified atmosphere has also been investigated.
In a similar vein, here we wish to study the interaction between two $\Omega$-loops but in a global spherical numerical setup that resembles that of the Sun
which includes convective motions, large scale flows and the Coriolis force.
We follow our study \cite{Jouve3} where $\Omega$-loops where introduced in a deep turbulent rotating convective
spherical envelope. The idea of this work is to identify configurations which will produce complex ARs. In particular, we will focus on the maps of the radial magnetic field emerging at the top of our computational domain, looking for non-bipolar (or multipolar) regions. Of strong interest are also of course the magnetic structures likely to produce eruptive events. This last property has been shown to be linked with the content of net electric currents in each polarity of the emerging region, \lau{as discussed in \citet{Forbes10} and \citet{2015ApJ...810...17D} and demonstrated in a recent observational survey \citep{Kontogiannis17}}. Indeed, these currents are good candidates to store the free magnetic energy necessary for flares and CMEs. The question then arises of how net currents can exist in each polarity where both direct and return currents are present \citep{1991ApJ...381..306M,1996ApJ...471..485P}.
Some MHD simulations have started to show that non-neutralized currents could be created through flux emergence \citep{2000ApJ...545.1089L,2014LRSP...11....3C} or photospheric flows \citep{Torok,2010ApJ...708..314A,2015ApJ...810...17D}. Through the 3D spherical simulations presented here where both emergence and horizontal flows are at play, we intend to identify the situations most likely to produce non-neutralized currents and thus the configurations most prone to solar eruptive events.\

 In section 2 we present the numerical setup of our parametric study and discuss a small analytical model that can
give us some clues on what to expect in the numerical simulations. In \S 3 we discuss the four configuration cases as defined in \cite{Linton}.
In \S4 we focus our attention on the resulting surface complexity of the emerging structure, while in \S 5 we develop a more objective criterion based 
on the computation of the radial current and conclude in section 6.

\section{Cases considered and expected interactions}

\subsection{Initial conditions to favor the creation of $\Omega$-loops}
\label{sect_init}

In this work, we investigate the different types of interactions between two magnetically buoyant structures rising through an unmagnetized convective or non-convective (or isentropic) environment. The initial conditions for the magnetic field are thus similar to the ones used in \cite{Jouve3} but we now initially introduce two axisymmetric flux tubes at the base of the convection zone. We here recall the expressions for these initial conditions. To ensure the solenoidal condition, a toroidal-poloidal decomposition is used:

\begin{equation}
{\bf B}={\bf \nabla\times\nabla\times}(C {\bf e}_r) +{\bf
\nabla\times}(A {\bf e}_r)
\end{equation}

\noindent where the expressions used for the potentials $A$ and $C$ are:

\begin{equation}
A=-A_{0} \, r \, \exp\left[-\left(\frac{r-R_t}{a}\right)^2\right]
\times
\left[1+\tanh\left(2\frac{\theta-\theta_t}{a/R_t}\right)\right]
\end{equation}

\begin{equation}
C=-A_{0} \frac{a^2}{2} \, q \,
\exp\left[-\left(\frac{r-R_t}{a}\right)^2\right] \times
\left[1+\tanh\left(2\frac{\theta-\theta_t}{a/R_t}\right)\right]
\end{equation}

\noindent where $A_0$ is a measure of the initial field strength, $a$ is the tube radius,
 $(R_t,\theta_t)$ is the position \lau{(radius and colatitude)} of the tube center and $q$ is the
 twist parameter.

\begin{figure}[h!]
\centering
\includegraphics[width=12cm]{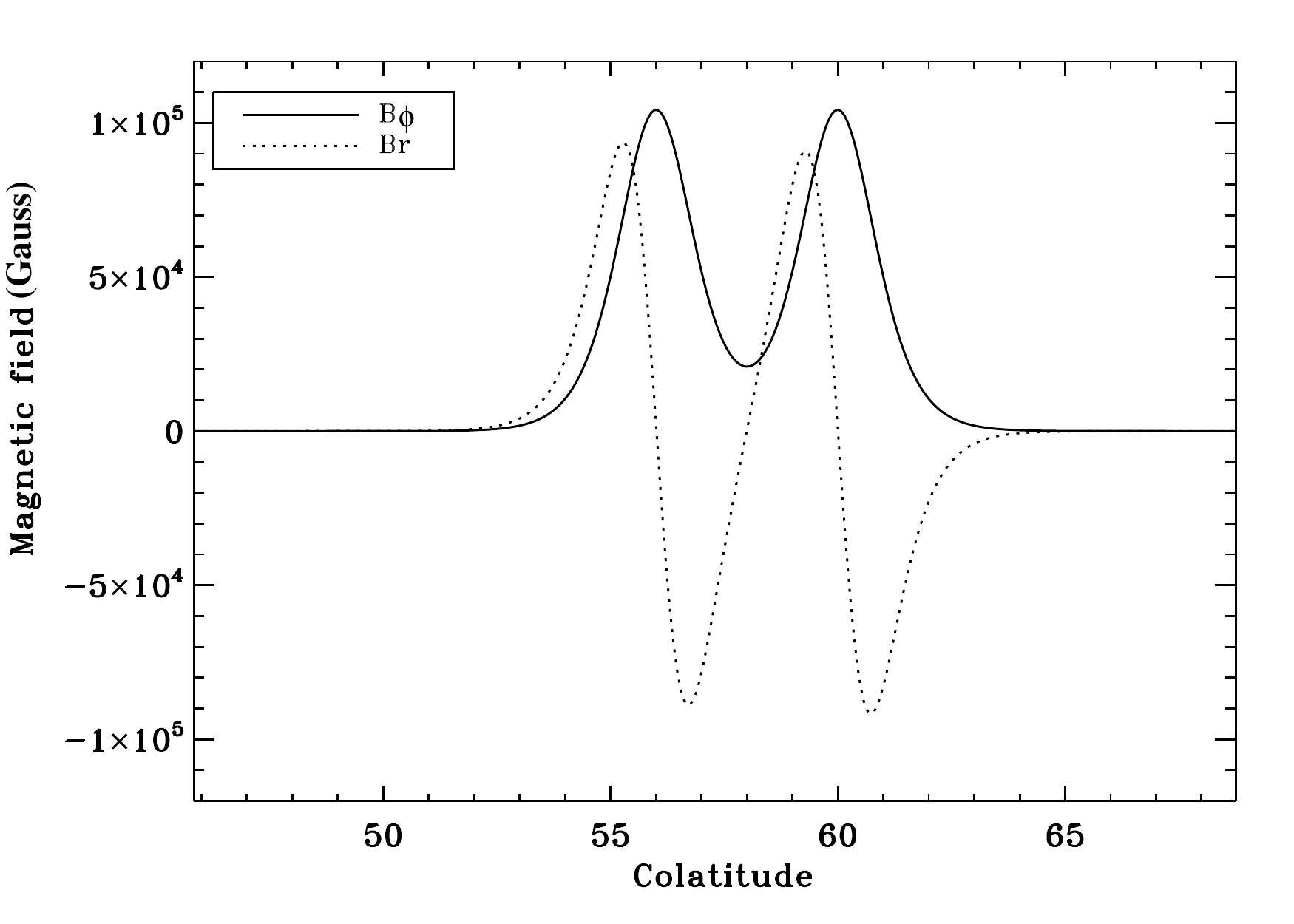}
\caption{Initial azimuthal ($B_\phi$) and radial ($B_r$) magnetic field as a function of colatitude, at radius $r=R_t$ for 2 loops initially separated by $4^o$ in latitude. Note that at this separation, the flux tubes are sufficiently close so that the azimuthal field does not completely vanish between the tubes. \lau{The peak toroidal field is located at the latitudes of introduction, i.e. $30^o$ and $34^o$, i.e. at $60^o$ and $56^o$ in colatitude.}}
\label{fig_init}
\end{figure}

 A typical initial condition for the radial and longitudinal components of the magnetic field of two loops located at colatitudes $\theta_{t1}=56^o$ and $\theta_{t2}=60^o$ \lau{(i.e. at latitudes $34^o$ and $30^o$, within the solar activity belt)}, at the same radius, with the same field strengths and twists and same tube radius is shown in Fig \ref{fig_init}. \lau{We refer to \citet{Jouve3} for a study of the effect of the initial latitude of a buoyant magnetic loop on its subsequent evolution.}

To have a measure of the degree of twist of the field lines in our cases, we can express the winding degree of the field
lines (i.e. the number of turns that the field lines make over the
whole tube length $2\pi R_t \sin \theta_t$) as: 

$$
n=\frac{\pi R_t \sin\theta_t}{2 a}\tan\psi
$$

 \noindent where the pitch angle $\psi$ is related to the twist parameter $q$ such that:
 
  $$
 \tan\psi=\frac{\sqrt{B_r^2+B_\theta^2}}{B_\phi}=q \frac{\sqrt{a^4+4(r-R_t)^2 r^2}}{2r^2}
 $$

In all cases, the tube radius for both structures will be
set to $a=2\times10^9 \,\, \rm cm$, about a tenth of the depth of the
modeled convection zone and will be introduced at
$R_t=5.2 \times 10^{10}\,\, \rm cm$, i.e. at around $0.75$ solar radius. \lau{We note that our main focus here is to study the interaction between magnetic structures which are sufficiently large-scale not to be strongly affected by magnetic diffusion, which is why we chose a relatively large radius of $20 \rm Mm$. However, this value is not unrealistic when dynamo simulations incorporating a subadiabatic layer are considered: for example, \citet{Browning06} do produce regions of strong toroidal field with an extent of about $30 \rm Mm$, i.e. larger than the tachocline thickness. We refer to Section 7.3 of \citet{2009ApJ...701.1300J} for a study of the effects of a buoyant loop radius on its evolution through the convection zone.} The initial field strength $A_0$, the initial twist of the field lines $q$ as well as the \lau{colatitude} of introduction $\theta_t$ of both flux tubes will be varied in our models.

In order to get a flux tube buoyant at particular longitudes only, we initially apply a perturbation on the background entropy field with a Gaussian profile in longitude $\phi$. The expression reads:

\begin{equation}
S^{\prime}=-A_{S} \times \exp\left[-\left(\frac{r-R_t}{a}\right)^2\right] \times
\frac{2R_t}{a} \times \frac{1}{\cosh^2 \left(2\frac{\theta-\theta_t}{a/R_t}\right)} \times
 \left[-C_S+\exp \left( -(\frac{\phi-\phi_0}{\phi_{ext}})^2\right) \right]
 \label{eq_s}
\end{equation}

 \noindent where $A_S$ is the amplitude of the entropy perturbation, $\phi_0$ and $\phi_{ext}$ are respectively the center and the extension of the entropy perturbation in longitude. This perturbation in entropy will produce a buoyant loop. The parameter $C_S$ enables to control the buoyancy of the regions outside this loop. \lau{It can be chosen so that the loop outside the buoyant part has the exact same density as the surroundings, to be in mechanical equilibrium.}

The effect of such a perturbation on entropy is to produce an additional density deficit inside the flux tube at a particular location in longitude, while the rest
of the flux tube can be maintained in mechanical equilibrium. As a consequence, we can derive the maximum density contrast between the tube and its surroundings, which will result from this perturbation.

\begin{equation}
\frac{\rho_{in}}{\rho_{ext}}=\exp\left(-\frac{Max(\vert S^{\prime} \vert )}{C_p}\right)\left(1-\frac{B^2}{8\pi P_{ext}}\right)^{1/\gamma}
\label{eq_rho}
\end{equation}

In the remaining of the paper, we will use the following parameters for the standard cases for the entropy perturbation: $C_S=-0.35$ and $\phi_{ext}=15^o$. \lau{We did not try to ensure strict mechanical equilibrium for the loop outside the buoyant part here. With our choice of $C_S$, the loop outside the buoyant part is actually slightly denser than the surroundings and will thus sink. The main results of the paper are however largely independent of this choice of $C_S$, as long as only a part of the loop is made buoyant.}

\subsection{Parameters for the different cases}

It has been shown previously in simpler configurations \citep[e.g.][]{Linton} that various types of reconnections could occur between flux tubes with different magnetic field strengths, field line twists and orientations. 

\begin{deluxetable*}{ccccccc}

\tablehead{
\colhead{\bf Case} & $\bf A_{01}$ & $\bf A_{02}$ &$\bf B_0$ & $\bf q_1$ & $\bf q_2$ &$ \bf n$ }
\startdata
CsTsB &   2000 & 2000 &  $10^5$ & 30  & 30 & 1.5\\ \hline
CsTwsB &   2000 & 2000 &  $10^5$ & 15  & 15 & 0.75\\ \hline
CsTsBw &   1200 & 1200 &  $6\times10^4$ & 30  & 30 & 1.5\\ \hline
CsTwsBw &   1200 & 1200 &  $6\times10^4$ &20  & 20 & 1\\ \hline
CoTsB &   2000 & 2000 &  $10^5$ & 30  & -30 & 1.5 \\ \hline
CsToB &   2000 & -2000 &  $10^5$ & 30  & 30  & 1.5\\ \hline
CoToB &   2000 & -2000 &  $10^5$ & 30  & -30  & 1.5\\ \hline
CoTsBw &   1200 & 1200 &  $6\times10^4$ & 30  & -30  & 1.5\\ \hline
CsToBw &   1200 & -1200 &  $6\times10^4$ & 30  & 30  & 1.5\\ \hline
CoToBw &  1200 & -1200 &  $6\times10^4$ & 30  & -30 & 1.5\\ \hline
\enddata

\caption{Summary of the cases with two $\Omega$-loops. $B_0$ indicates the initial axial field strength in Gauss and $n$ indicates the number of turns of the field lines along the $15^o$ longitudinal extension.}
\label{tab}
\end{deluxetable*}

The loops are initially introduced at neighbouring colatitudes \lau{within the latitudinal belt of solar activity} (Loop 1 at $\theta_{t1}=60^o$, \lau{or $30^o$ latitude} and Loop 2 at $\theta_{t2}=56^o$ \lau{or $34^o$ latitude}) and at the same radius $R_t$. This choice is motivated by previous 2D and 3D numerical studies of magnetic buoyancy instabilities of a magnetic layer \citep[e.g.][]{Cattaneo88, Matthews95, Wissink}, which show that 2D interchange modes first tend to be unstable to a Rayleigh-Taylor instability. The most unstable modes possess high wave numbers, especially when low values of the magnetic diffusivity are considered. This unstable situation thus gives rise to flux tubes located very close to each other and which can then undergo a secondary 3D instability, producing arched structures. We thus intend to explore here the interactions of two such neighboring structures in a spherical shell.

 The initial twist and direction of the field lines and the buoyancy of each loop are then varied. For the latter, we can play on two different parameters: either the field strength or the entropy perturbation since they both control the buoyancy of the flux tube (see eq. \ref{eq_rho}). We choose to focus on the effect of the magnetic field strength and thus fix the amplitude of the entropy perturbations to $A_{S1}=200$ and $A_{S2}=150$. In the cases considered, all loops will rise through the convection zone in approximately 8 to 10 days, Loop 1 being potentially slightly faster than Loop 2. \lau{From eqs. \ref{eq_s} and \ref{eq_rho}, adopting the values $C_p=3.4\times10^8 \,\, \rm erg.K^{-1}.g^{-1}$, $P_{ext}= 4\times 10^{13} \,\, \rm dynes$ and $\gamma=5/3$, we can calculate a relative density deficit in the loops: $\Delta \rho/\rho$ goes from $1.7\times 10^{-5}$ for the less buoyant case ($B_0=6\times 10^4 \rm G$ and $A_S=150$) to $2.6\times 10^{-5}$ in the most buoyant case ($B_0=10^5 \rm G$ and $A_S=200$) (see also Fig.6 of \citet{2013ApJ...772...55P}). These relatively small values are comparable in amplitude to the fluctuations measured at the base of our computational domain in the convective simulations.} The maximum entropy perturbation will always be located at $\phi_{01}=100^o$ for Loop 1 and $\phi_{02}=92^o$ for Loop 2, to represent neighboring flux tubes whose field lines are bent in the direction of the axial field, but at different (though neighboring) positions in longitude. 
 
 \lau{We wish to initialize our simulations with values of the magnetic field strength at the bottom of the convection zone of the order of a few times the equipartition field (same energy as the strongest down flows at the base of the convection zone) and a reasonable degree of twist based on active region observations \lau{(\citet{Chae05} for example quote a winding number of less than 0.75 over the whole active region.)}. For these reasons, we chose typical field strengths between $50$ and $100 \rm kG$ (compatible with the values of the buoyant field produced in the 3D dynamo simulations of \citet{2014SoPh..289..441N} or \citet{2014ApJ...789...35F} for example), twist parameters between $15$ and $30$, corresponding to $0.75$ to $1.5$ turns along the loop extension of $\phi_{ext}=15^o$.}

Table \ref{tab} summarizes the various cases which were computed and the main parameters used. \lau{The letter `C' stands for `Case'. Capital `T' stands for `Twist' and `B' for the axial magnetic field direction. Cases with the same twist are labeled with the letters `sT' and with opposite twist with `oT'. Identically, the cases with the same axial field orientation are labeled with `sB' and with opposite orientation with `oB'. When the letter `w' is added after `T' or after `B', it indicates a case with a weak twist or a weak field. For example, CsTwsB corresponds to a case with the same weak value for the twist, and the same axial field direction.}

\begin{figure}[h!]
\centering
\includegraphics[width=16cm]{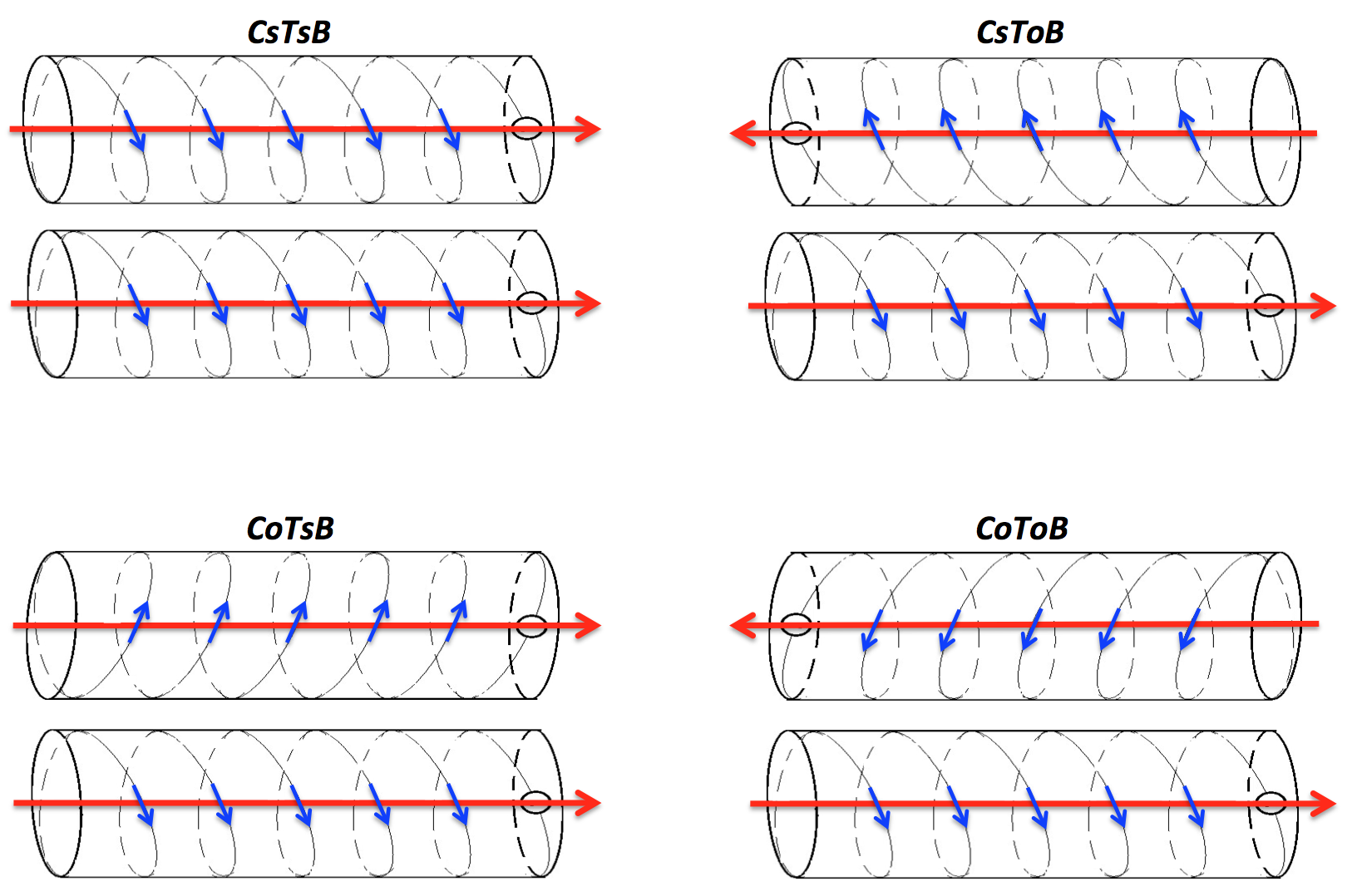}
\caption{Schematic representation of the different field line directions in a portion of the initial flux tubes in the 4 cases considered in this work. \lau{The red arrow indicates the direction of the axial field and the blue arrows that of the twist of the field lines.}}
\label{fig0}
\end{figure}

The first 4 cases CsTsB, CsTwsB, CsTsBw and CsTwsBw are computed to investigate the impact of magnetic field strength and twist intensity on the reconnection which could occur between loops of similar orientation, field strength and twist. The other cases CoTsB, CsToB, CoToB and their equivalent with a weaker initial field CoTsBw, CsToBw and CoToBw are dedicated to the study of the effects of different field line orientations. Fig.\ref{fig0} shows a schematic representation of the various initial field configurations considered for the two loops.

\subsection{Expected interactions between loops}
\label{sect_localc}

From previous work, such as \cite{Linton} performed in a local Cartesian geometry without convection, we have an idea of the possible interactions between our loops possessing different field line orientations. When the magnetic field along the loop axis and the twist of the field lines are of the same sign (CsTsB), the loops are expected to undergo reconnections since the field lines getting close to each other at the loop peripheries will be anti-parallel to each other. When the twist parameters or the field along the axis are of opposite signs (CoTsB or CsToB), then reconnection is not expected since the field lines are parallel to each other. Finally when both the twist parameter and axial field are of opposite signs in the two loops (CoToB), then full reconnection is expected, producing a large energy conversion from magnetic to kinetic. These different interactions were also studied in the same kind of setup by \cite{Sakai92} and in a more realistic situation mimicking reconnecting coronal loops by \cite{Ozaki97}. We here notice that we will only consider cases with two interacting loops, instead of one loop with two buoyant regions, as was studied by \citet{2014SoPh..289.3351T}. We did analyse few simulations of such setups but the buoyant regions did not strongly interact before reaching the top of our domain, they will however interact after flattening against the surface (as shown in \citet{2014SoPh..289.3351T}) but we do not intend to model this late evolution here. This situation is thus of little interest if we consider the emergence through the whole convection zone.

To get more insight into the expected interactions between our loops, we initially calculate the Lorentz force caused by the presence of both loops. We focus on its latitudinal component since it will act on the $\theta$-component of the velocity which will then advect the loops towards each other (attractive force) or on the contrary away from each other (repulsive force).

We thus calculate the latitudinal component of the initial Lorentz force, namely:

\begin{equation}
F_\theta=\frac{1}{c}{\bf J \times B} \vert_{e_\theta}=\frac{1}{c}(J_\phi  B_r - B_\phi  J_r)
\end{equation}

with ${\bf J}=c/4\pi \,\, {\bf \nabla \times B}$ where $c$ is the speed of light.

\begin{figure}[h!]
\centering
\includegraphics[width=18cm]{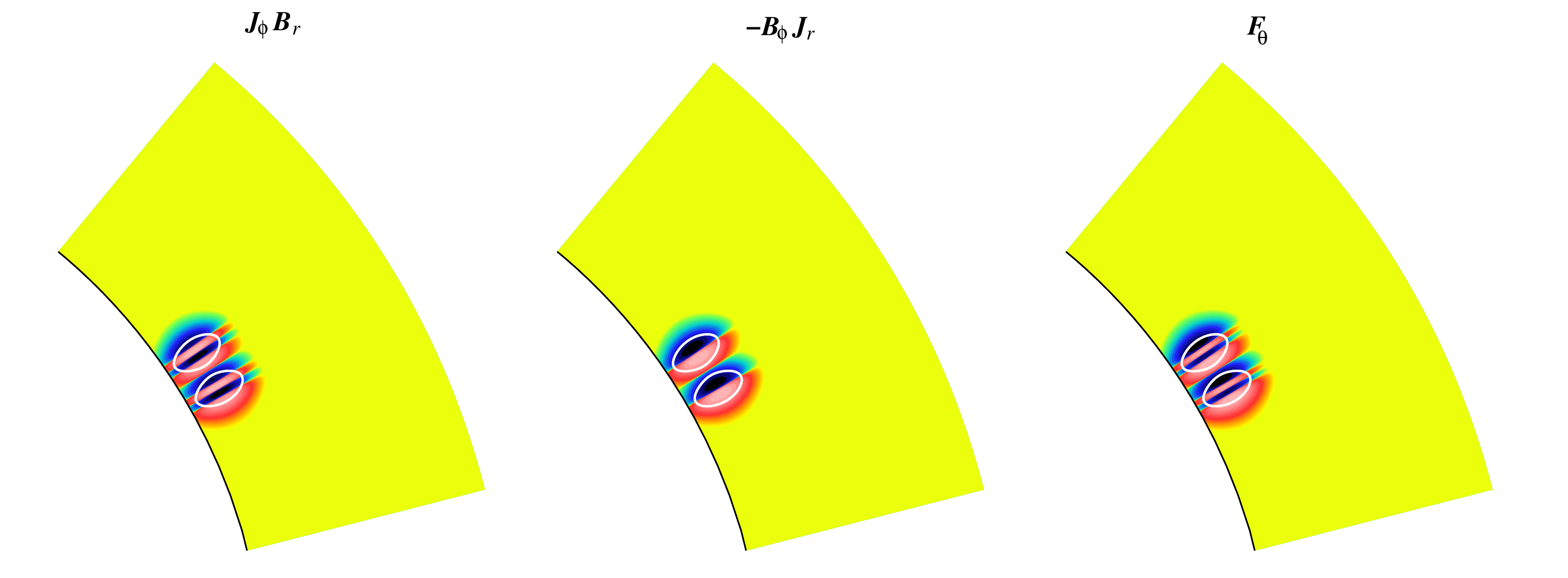}
\caption{Terms $J_\phi  B_r$ (left) and $- B_\phi  J_r$ (middle) adding up to give the latitudinal Lorentz force $F_\theta$ (right) in the meridian plane, for CsTsB. \lau{Red (resp. blue) color indicates positive (resp. negative) values and the white circles are the contours of the initial toroidal field corresponding to an amplitude of $3\times10^4 \rm G$. Near the contact point between the loops, a positive region of $F_\theta$ lies over a negative region, indicating a net attractive force between the loops.}}
\label{fig1}
\end{figure}

\begin{figure}[h!]
\centering
\includegraphics[width=12cm]{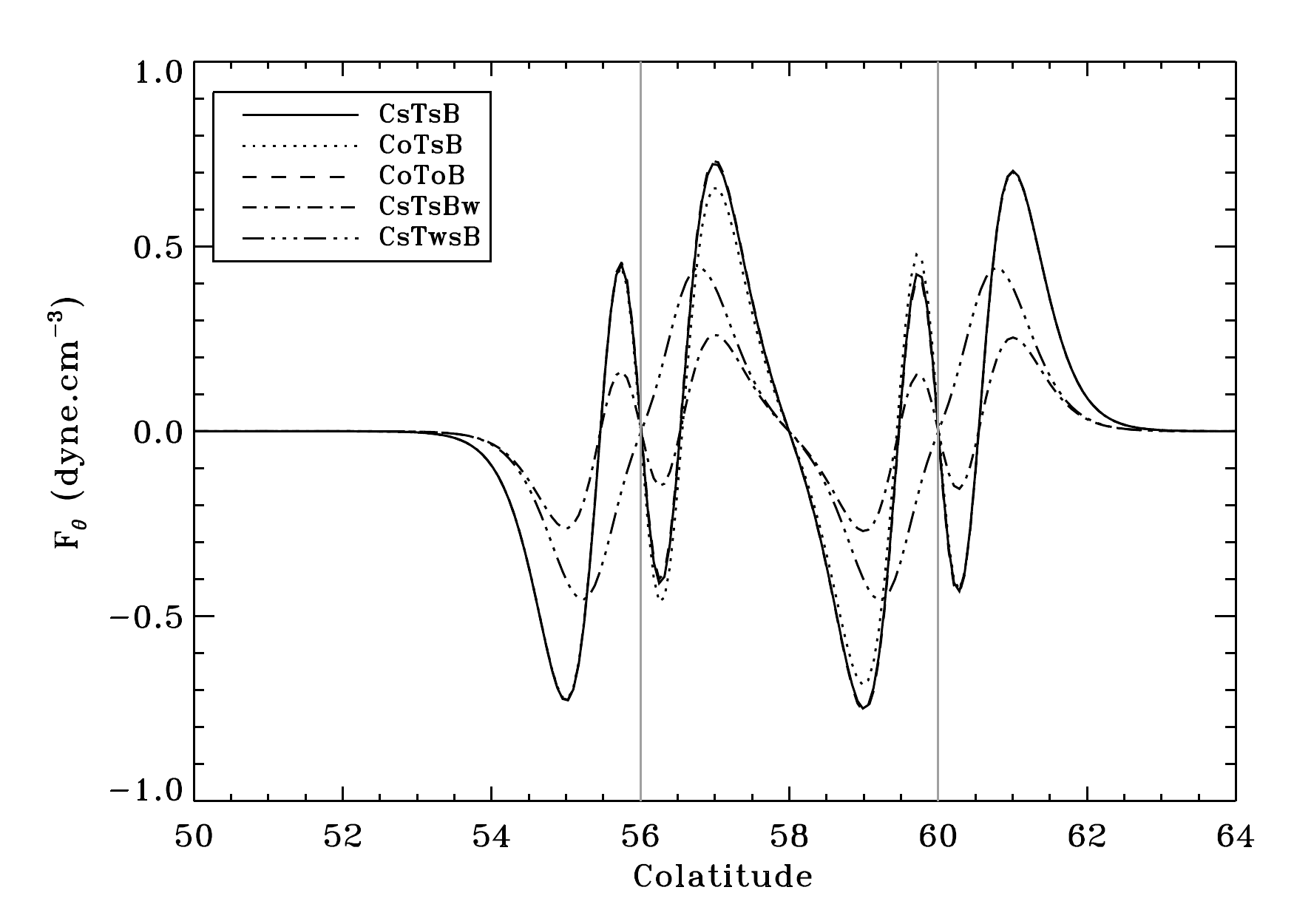}
\caption{Latitudinal Lorentz force per unit volume, as a function of colatitude (in degrees), at radius $r=R_t$, for an initial angular separation of $4^o$ in Cases CsTsB, CoTsB, CoToB, CsTsBw and CsTwsB. The two vertical grey lines indicate the positions in colatitude of the axis of the two loops, \lau{corresponding to latitudes $30^o$ and $34^o$. Between those lines, the Lorentz force is attractive.}}
\label{fig2}
\end{figure}

Figures \ref{fig1} and \ref{fig2} show the initial latitudinal Lorentz force calculated for different cases shown on Table \ref{tab}. Figure \ref{fig1} shows the distribution of the two terms involved in the calculation of $F_\theta$, namely  $J_\phi  B_r$ and $- B_\phi  J_r$ in a meridian plane. The first term will be sensitive to a change of the twist parameter while the second one will be modified when the initial field strength of the tube is changed. This calculation was done for the parameters of CsTsB but changing the sign of the twist or the azimuthal field orientation in one loop does not modify at all the values of these different terms when the loops are sufficiently distant. Indeed, when there is only little overlap between the magnetic field concentrations, changing the sign of $B_r$ in one tube changes the sign of $J_\phi$ and thus the end product will remain the same (this is also true for the other term $- B_\phi  J_r$). As a consequence, in all the cases we consider \lau{with a separation of $4^o$}, the latitudinal Lorentz force is attractive between the loops. If we focus on Figure \ref{fig2}, we see that between colatitudes $57^o$ and $59^o$, i.e. exactly in the middle of the two loops, $F_\theta$ goes from positive to negative, meaning that at higher latitudes it is oriented towards the equator and at lower latitudes towards the pole. We note again that for all cases here with an initial latitudinal separation of $4^o$, $F_\theta$ is independent of the field line orientations. Of course when the initial field strength is reduced (corresponding to CsTsBw), $F_\theta$ is reduced by the same amount and the attractive force between the loops is smaller. It is also interesting to note that when the initial twist of the field lines is reduced (CsTwsB), the intensity of the term $J_\phi  B_r$ is reduced compared to $B_\phi  J_r$. As a consequence, the term $B_\phi  J_r$ can become dominant. The sign of this term remains constant over a larger extent in latitude and this leads to a latitudinal Lorentz force which will be attractive on a more extended region in latitude, as seen in Fig. \ref{fig2} for CsTwsB.

\begin{figure}[h!]
\centering
\includegraphics[width=8.5cm]{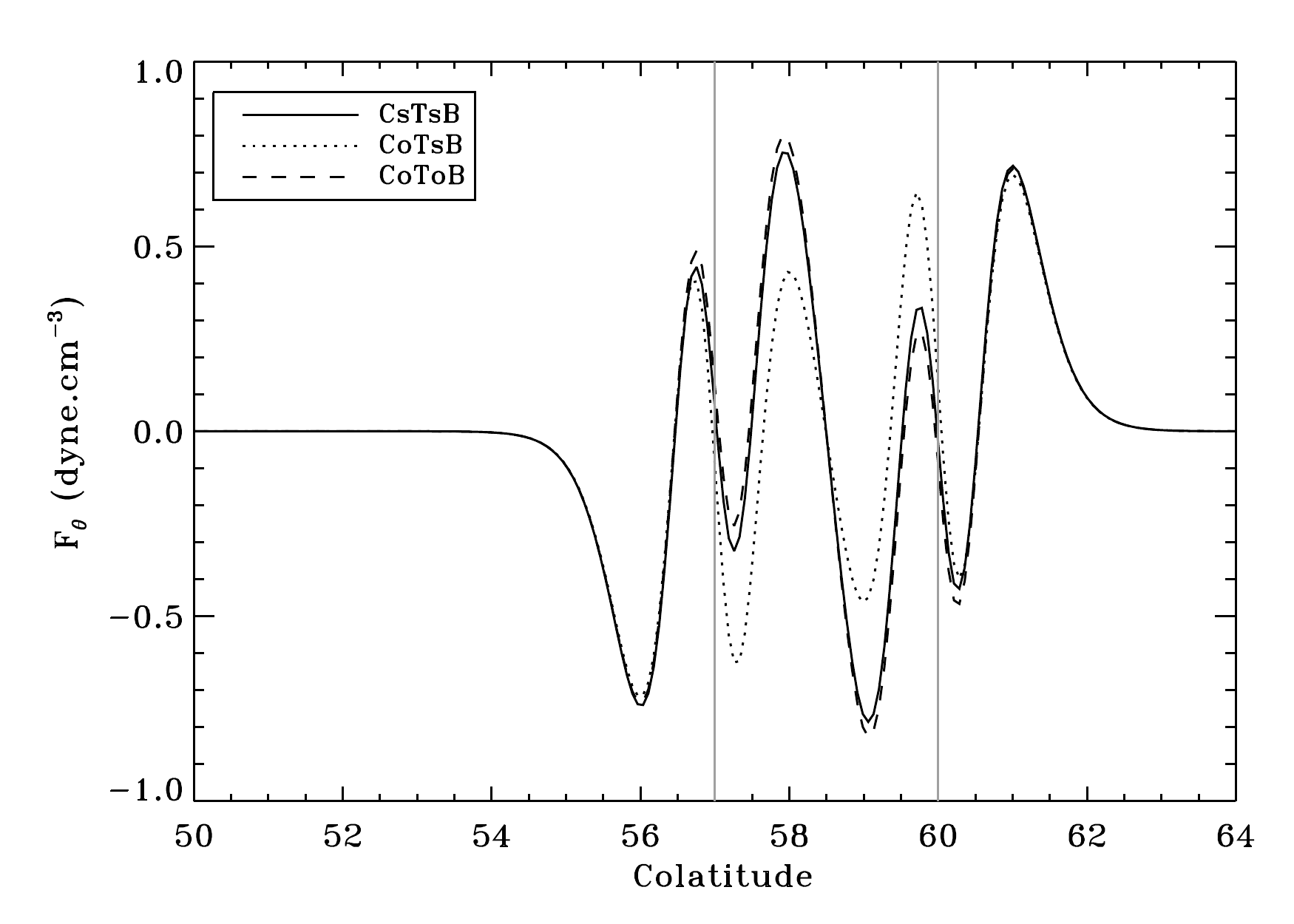}
\includegraphics[width=8.5cm]{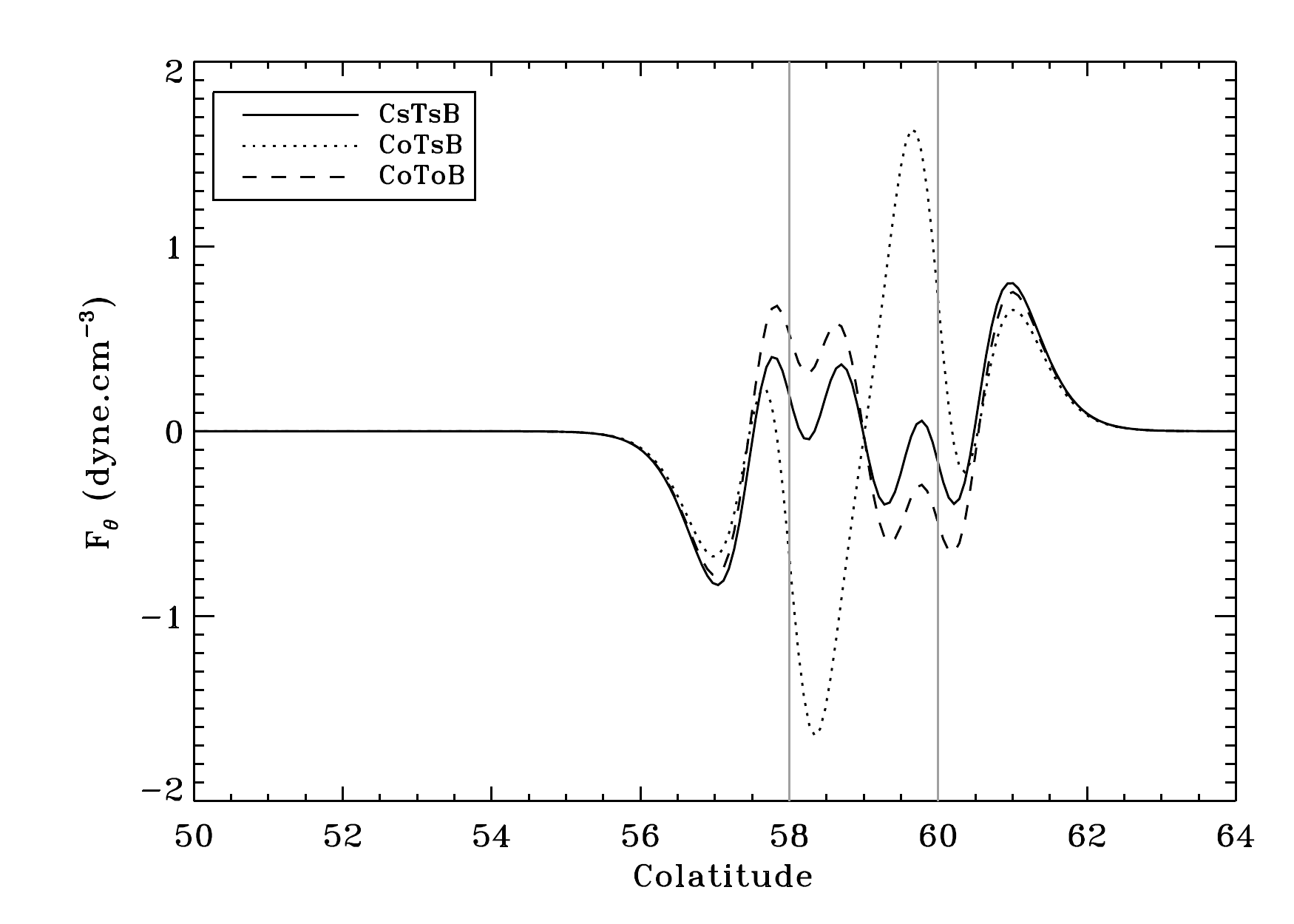}
\caption{Same as Fig.\ref{fig2} (without CsTsBw which is similar to CsTsB but only with a smaller amplitude) but for smaller angular separations: $3^o$ on the left panel \lau{(the grey lines correspond to the latitudes $30^o$ and $33^o$)} and $2^o$ in the right panel \lau{(the grey lines correspond to latitudes $30^o$ and $32^o$). On the right panel, note that in CoTsB, the Lorentz force between the grey lines is now repulsive.}}
\label{fig3}
\end{figure}

Since these loops initially exert an attractive force on each other, the natural evolution for them will be to get closer to each other. We can thus recalculate the Lorentz force in our different cases, assuming that the magnetic field is still equal to its initial expression. This will presumably not be the case in our full 3D calculations, but reducing the latitudinal separation between the loops in our analytical calculations will help to gain some insight on the expected interactions between our structures when they start to significantly overlap.
When the initial separation is decreased to $3^o$ and then $2^o$, the distinction between the different cases starts to be more obvious. This is what is shown in Figure \ref{fig3}: the latitudinal Lorentz force $F_\theta$ is shown as a function of the colatitude for cases CsTsB, CoTsB and CoToB with an initial separation of $3^o$ (left panel) and $2^o$ (right panel). 

For $3^o$ of separation, $F_\theta$ remains attractive between the loops for all loops but is strongly decreased in the ``opposite twist" case (CoTsB). This implies that the loops should continue to be advected towards each other and we thus move to an initial separation of only $2^o$. In this case, the difference between the ``same twist" and ``opposite twist" cases is drastic. The latitudinal Lorentz force between the loops in CoTsB now becomes strongly repulsive, which implies that the loops will now have the tendency to be pushed away from each other. The attractive force in CsTsB is also strongly reduced but still remains quite strong in CoToB. Strong interactions are thus expected in this case where the loops will be continuously attracted to each other until the magnetic field of opposite polarities cancel each other.  

These analytical results, although limited since they do not take into account the dynamical evolution of the magnetic field, are in good qualitative agreement with the previous results of \cite{Linton} or \cite{Sakai92} and guide us towards interesting choices of parameters. Firstly, a latitudinal separation of more than $4^o$ may strongly limit the possible interactions between the loops since they may not feel the initial attractive force that they exert on each other. \lau{We confirmed that using a separation of $5^o$ in the 3D simulations ends up in almost no interaction between the loops}. We will thus choose an initial separation of $4^o$ so that the loops will be initially naturally pushed towards each other. \lau{We note that this separation has to be decreased for thinner tubes to get this initial attraction since the peaks of the Lorentz force will be more localized. This is of course neglecting the possible effects of convective motions or mean flows which could advect the magnetic structures towards each other. Secondly, the initial magnetic field amplitude and the initial field line twist directly impact the value of the initial $F_\theta$ and from this analytical study, we know approximately which initial field amplitude and twist are needed for the attractive Lorentz force to be effective. These are compatible with the parameter values listed in table \ref{tab}.}
%We also wish to initialize our simulations with realistic values of the magnetic field strength at the bottom of the convection zone and a reasonable degree of twist based on active regions observations. For these reasons, we chose typical field strengths between $50$ and $100 \rm kG$, twist parameters between $15$ and $30$, corresponding to $0.75$ to $1.5$ turns along the loop extension of $\phi_{ext}=15^o$ and an 

 We now want to confront these analytical results with fully non-linear 3D numerical simulations of interacting loops in a convective rotating spherical shell. The major differences here are that the dynamical evolution of the magnetic structures and their non-axisymmetry are now accounted for and that our structures will be embedded in convective motions to mimic the conditions of the solar convective zone. This of course may significantly change those results since an additional latitudinal velocity due to convective motions may compete with that induced by the latitudinal Lorentz force calculated in this section.

\section{Full 3D numerical simulations of interacting loops}
\subsection{Anelastic MHD equations and hydrodynamical background}

The simulations described below were performed with the anelastic
spherical harmonic (ASH) code, which solves the three-dimensional
anelastic equations of motion in a rotating spherical shell using a
pseudo-spectral semi-implicit approach
\citep[e.g.][]{Miesch0,Brun2}. The equations
are fully nonlinear in velocity and magnetic fields and linearised in
thermodynamic variables with respect to a spherically symmetric mean
state to have density $\bar{\rho}$, pressure $\bar{P}$, temperature
$\bar{T}$, specific entropy $\bar{S}$. Perturbations are denoted as
$\rho$, $P$, $T$ and $S$. The equations being solved are

\begin{equation}
{\bf\nabla}\cdot(\bar{\rho}{\bf v})=0,
\end{equation}
 \begin{equation} {\bf\nabla}\cdot{\bf B}=0,
\end{equation}
\begin{eqnarray}
\bar{\rho}[\frac{\partial{\bf v}}{\partial t}+({\bf v}\cdot{\bf
    \nabla}){\bf v}&+&2\Omega_0\times {\bf v} ]=-{\bf \nabla} P
+\rho{\bf g}\\ \nonumber
&+&\frac{1}{4\pi}{\bf (\nabla \times B) \times B}
-{\bf \nabla \cdot \cal D}-[{\bf \nabla}\bar{P}-\bar{\rho}{\bf g}],
\label{eqNS}
\end{eqnarray}
\begin{eqnarray}
& &\bar{\rho} \bar{T}\frac{\partial S}{\partial t}+\bar{\rho}\bar{T}{\bf
  v}\cdot{\bf \nabla}(\bar{S}+S)={\bf
  \nabla}\cdot [\kappa_r\bar{\rho}c_p{\bf
    \nabla} (\bar{T}+T)\\ \nonumber
&+&\kappa_{0}\bar{\rho}\bar{T}{\bf
    \nabla}\bar{S}+\kappa \bar{\rho} \bar{T}{\bf \nabla} S ]+\frac{4\pi\eta}{c^2}{\bf j}^2 
+2\bar{\rho}\nu\left[e_{ij}e_{ij}-\frac{1}{3}({\bf \nabla \cdot
v)}^2\right],
\end{eqnarray}
\begin{equation}
\frac{\partial {\bf B}}{\partial t}={\bf \nabla \times} ({\bf v\times \bf B})-{\bf \nabla \times}(\eta {\bf \nabla \times B})
\end{equation} 

\noindent where ${\bf v}=(v_r,v_{\theta},v_{\phi})$ is the local
velocity in spherical coordinates in the frame rotating at a constant
angular velocity $\Omega_{0}$, ${\bf g}$ is the gravitational
acceleration, ${\bf B}=(B_r,B_{\theta},B_{\phi})$ is the magnetic
field, ${\bf j}=(c/4\pi)({\bf \nabla \times B})$ is the current
density, $c_p$ is the specific heat at constant pressure, $\kappa_r$
is the radiative diffusivity, $\eta$ is the effective magnetic
diffusivity and $\cal D$ is the viscous stress tensor. The coefficients $\nu$ and $\kappa$ are assumed to be an effective eddy viscosity and
eddy diffusivity, respectively, that represent unresolved
subgrid-scale processes, chosen to accommodate the
resolution. \lau{The effective diffusivities are all chosen to vary as $\bar{\rho}^{-1/3}$ in the present simulations, in particular, the magnetic diffusivity $\eta$ varies from $8\times10^{11}\rm cm^2.s^{-1}$ at the bottom of the convection zone to $2.3\times10^{12} \rm cm^2.s^{-1}$ at the top}. The thermal diffusion $\kappa_0$ acting on the mean
entropy gradient occupies a narrow region in the upper convection
zone. Its purpose is to transport heat through the outer surface where
radial convective motions vanish \citep{Gilman81, Wong94}. To complete the set of equations, we use the
linearised equation of state

\begin{equation}
\frac{\rho}{\bar{\rho}}=\frac{P}{\bar{P}}-\frac{T}{\bar{T}}=\frac{P}{\gamma\bar{P}}-\frac{S}{c_p}
\end{equation}

\noindent where $\gamma$ is the adiabatic exponent, and assume the ideal gas law

\begin{equation}
\bar{P}={\cal R} \bar{\rho}\bar{T}
\end{equation}

\noindent where $\cal R$ is the ideal gas constant, taking into
  account the mean molecular weight $\mu$ with 3/4 of Hydrogen and 1/4 of Helium per mass. The reference or
mean state (indicated by overbars) is derived from a one-dimensional
solar structure model (obtained by the 1D CESAM stellar evolution code \citep{Morel97})
and is regularly updated with the spherically
symmetric components of the thermodynamic fluctuations as the
simulation proceeds \citep{Brun2002}.

The computational domain extends
from about $0.72 R_{\odot}$ to $0.96 R_\odot$. At those boundaries, the velocity is impenetrable and
stress-free. We impose a constant entropy gradient top and bottom for the isentropic case and for the
fully convective case, a latitudinal entropy gradient \lau{(corresponding to a temperature difference between pole and equator of about $10 \rm K$)} is
imposed at the bottom, as in \cite{Miesch}, to mimic a solar-like differential rotation. \lau{The differential rotation profile is similar to the one established in \citet{2009ApJ...701.1300J}: the angular velocity contours are radial at mid-latitudes and the rotation period goes from 25 days at the equator to about 35 days close to the poles, in agreement with helioseismic inversions \citep{Thompson03}.}
In all cases, we match the magnetic field to an external potential field at the top and the bottom of the shell \citep{Brun2}.

Our experiments consist in introducing torii of magnetic field at
the base of the convection zone in a spherical shell, in a thermally
equilibrated hydrodynamical model in which the convection is or is not triggered. The hydrodynamical background models are the same as the ones used in \cite{Jouve3}, 
we thus refer to this article for the exact values of the parameters. We only recall that in all cases, the density
contrast is about 24 between the top and the bottom of
the domain, that the Prandtl number is $P_r=0.25$ and that in the magnetic cases, the magnetic Prandtl number is set to $P_m=1$.
In the non-convective (or isentropic) cases, the spherical shell rotates rigidly at the solar rotation rate $\Omega_0=2.6 \times 10^{-6} \rm rad.s^{-1}$ 
whereas in the convective case, a differential rotation naturally develops, with a fast equator and slow poles. The convective model also possess a large-scale meridional circulation, organized as one large cell per hemisphere, directed poleward at the surface, with a maximum amplitude of about $20 \rm m.s^{-1}$. The typical rms velocity of the convective flows reach values of approximately $200 \rm m.s^{-1}$ in the bulk of the convection zone.

\subsection{Temporal evolution of energies in the different cases}
\label{sect_energies}

Before discussing the details of the evolution of our magnetic structures, we investigate the global properties of our simulations and in particular the temporal evolution of the kinetic and magnetic energies. As will be discussed in more details in the following sections, the interactions that occur between the two loops agree with the expectations of Sect. \ref{sect_localc}. Indeed, CsTsB where the magnetic field has the same orientation in both loops is a case where the two loops are likely to merge while they rise through the convection zone. Cases CoTsB and CsToB where either the twist or the axial field in both loops are opposite do not undergo such a merging. Finally, CoToB where both the axial field and the twist are opposite is yet another situation, where all components of the magnetic field are likely to reconnect. Since the loops are introduced as axisymmetric flux tubes and that the environment is convective, it is misleading to consider the energies of the whole spherical shell. We thus concentrate on the maxima of kinetic and magnetic energies in the area where the buoyant structures are confined.

\begin{figure*}[h!]
	\centering
	\includegraphics[width=8.5cm]{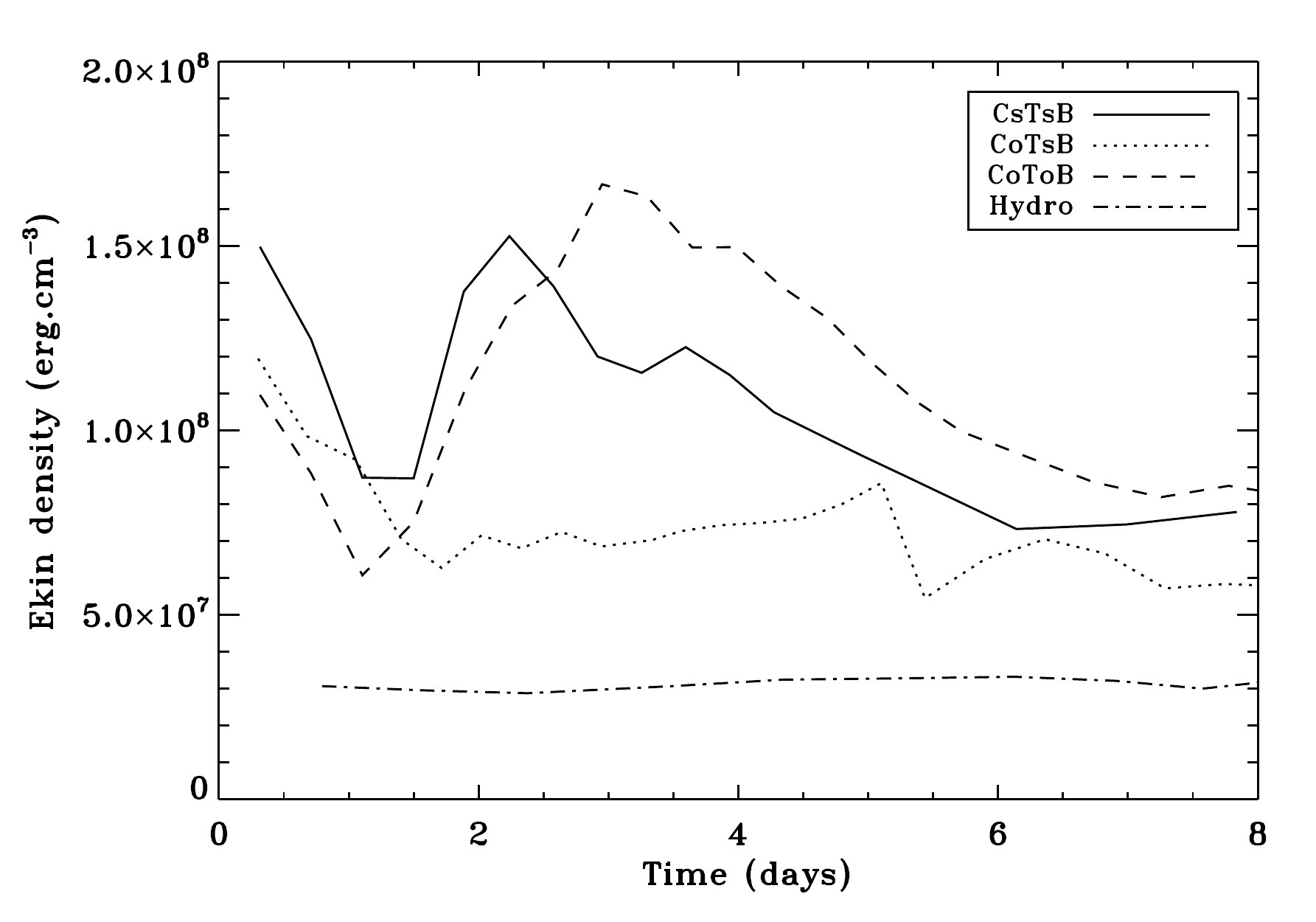}
	\includegraphics[width=8.5cm]{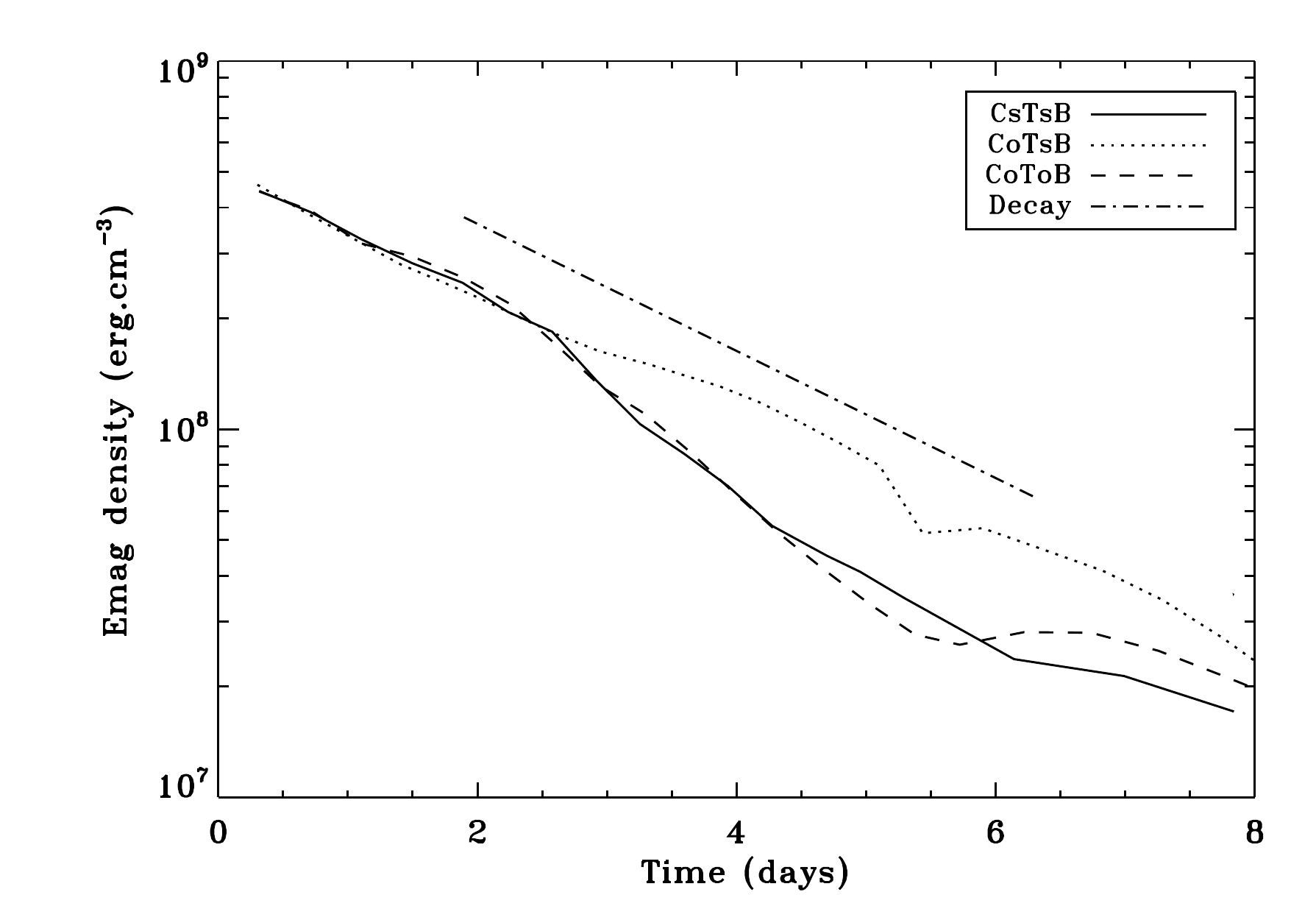}
	\caption{Temporal evolution of the maximum of kinetic and magnetic energies in the confined area where the loops evolve for Cases CsTsB, CoTsB and CoToB. The purely hydrodynamical case is also added in the kinetic energy plot for comparison \lau{and the constant slope corresponding to a purely Ohmic diffusion is added on the magnetic energy, labelled as ``decay''. A conversion of magnetic to kinetic energies is clearly visible for the merging cases CsTsB and CoToB.}}
	\label{fig_energies}
\end{figure*}

The temporal evolution of these maxima of kinetic and magnetic energies is plotted in Figure \ref{fig_energies} for cases CsTsB, CoTsB and CoToB for which the types of interactions between the loops are different. For the kinetic energy, we also show for comparison the value of its maximum in the same area when no magnetic field was introduced (hydro case).

These plots allow us to clearly distinguish the cases where an interaction between the loops occurs and the case where this interaction is absent. If we look at Cases CsTsB and CoToB, a large sudden increase of kinetic energy (KE) appears at around 2 to 3 days of evolution, associated with a decrease of magnetic energy (ME). As we shall see later, it is exactly at that time that the loops start to merge in Cases CsTsB and CoToB. As we can see in Fig. \ref{fig_energies}, the increase in KE is not exactly coincidental with the dip in the ME curve. This is due to the fact that KE first increases because of the increase of the Lorentz force when the loops gets advected towards each other because of the attractive force discussed in the previous section. When the loops are then sufficiently close, reconnection of the field lines occurs at the periphery of the structures. The topology of the field lines and in particular the distribution of their helicity changes, constituting an efficient mechanism of conversion from magnetic to kinetic energies. We note that the large increase in KE is more pronounced in CoToB where reconnection of the field lines is likely to happen at the loop peripheries but also along the tube axis since the axial magnetic fields are anti-parallel in this case. On the contrary, for \lau{CoTsB} where the loops only bounce against each other because of the repulsive Lorentz force they exert (see previous section), the KE curve remains flat and the ME decreases solely because of magnetic diffusion. \lau{For comparison, an exponentially decaying function of time is superimposed on the magnetic energy plot (labelled ``decay") with a characteristic timescale $\tau=l^2/\eta$ with $l=10^9\rm cm$ the typical length scale of our loops and $\eta=2.3\times10^{12} \rm cm^2.s^{-1}$ the value of the magnetic diffusivity at the top of our domain. We recover that the magnetic energy in case CoTsB decays at this same expected rate.}

\subsection{Same direction of axial field and twist: attraction and merging}

This situation is thought to be the most plausible during a particular magnetic cycle since both loops have the same sign for the axial field and for the twist of the field lines. We indeed expect that if buoyancy instabilities produce the flux ropes which then rise to the surface, they will be triggered on a strong toroidal field of a particular sign in a given latitudinal band in each hemisphere. The twist of the field lines is also thought to possess a preferred sign in each hemisphere. In this situation, the loops are likely to reconnect at their periphery if they can get close enough to each other.

\subsubsection{Typical case of merging loops}

\begin{figure*}[h!]
	\centering
	\includegraphics[width=17cm]{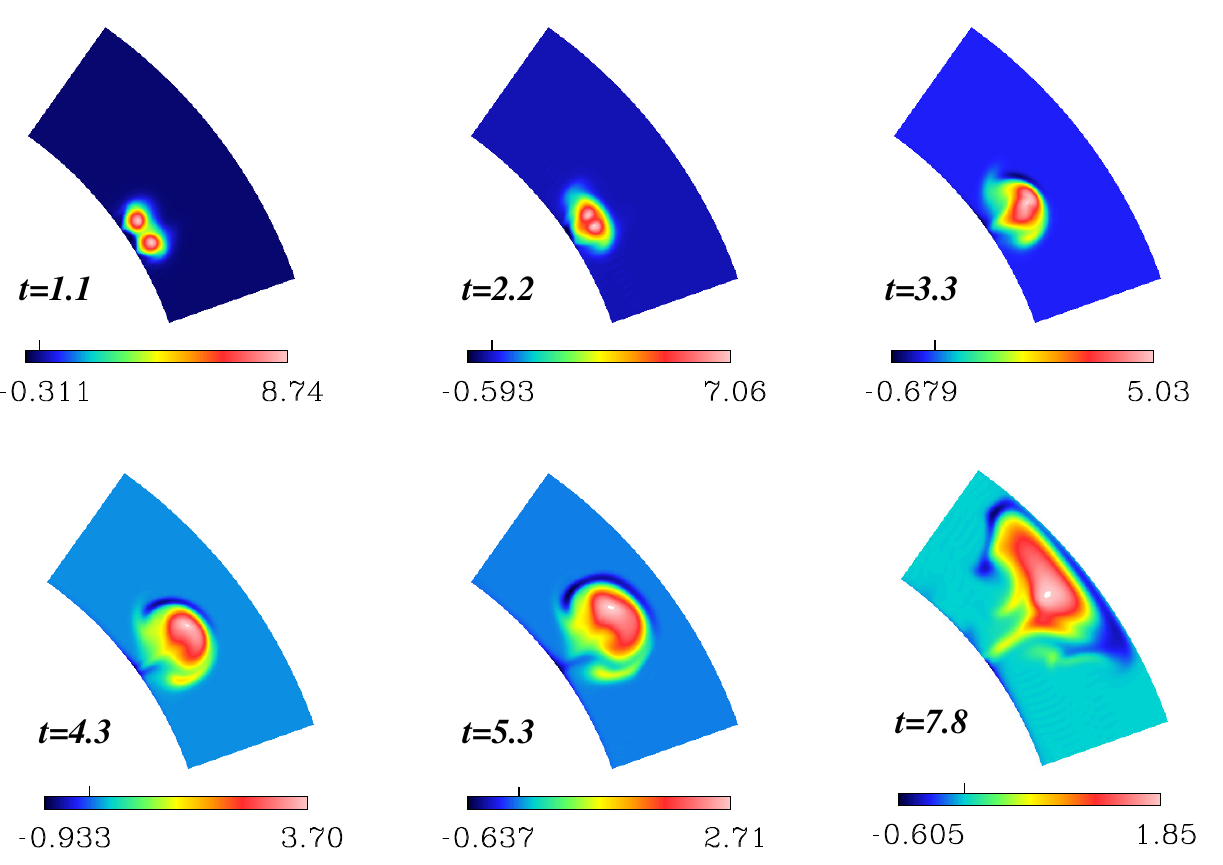}
	\caption{CsTsB: cut at the particular longitude of $90^o$ of the toroidal magnetic field during the evolution of the loops through the convection zone. \lau{The values are in $10^4 \rm G$ and on the color bars, the small vertical line indicates the color corresponding to $0$. In this case, the loops merge after about 3 days of evolution to produce one single emerging structure.}}
	\label{fig_merge}
\end{figure*}

CsTsB is a typical case of such a situation where the loops merge after around 3 days of evolution through the convection zone. We recall that the loops are here initially introduced with a magnetic field intensity of about $10^5 \rm G$, with the same orientation of the axial field lines and with a twist of the same sign. This implies that the field lines facing each other at the loops peripheries are anti-parallel and thus likely to reconnect. We also recall that an initial separation of $4^o$ in latitude is chosen so that the loops are initially attracted to each other, as discussed in Section \ref{sect_localc}. Figure \ref{fig_merge} shows a cut in the meridian plane of the toroidal magnetic field contained in the loops, from the first day of evolution to the time at which the magnetic concentration reaches the top of our domain. 

From the starting time to around 3 days, the evolution consists in the loops getting dragged towards each other because of the attractive Lorentz force discussed in the first section. We did verify that the amplitude and profile of the latitudinal Lorentz force calculated at the beginning of the simulation is similar to the analytical predictions. At time $t=2.22$ days, the loops start to get very close to each other and the increase of kinetic energy seen in Fig.\ref{fig_energies} has already started. At around $t=3$ days, a reconnection of the field lines occurs, producing a topological change of the magnetic configuration. The magnetic concentration now consists in a unique flux tube with a total toroidal and poloidal field with a reduced intensity: at time $t=4.27$ days, the maximum magnetic field contained in the loop is only about $37\%$ of its initial value in each loop. This is consistent with the dip in the magnetic energy which was observed in Fig.\ref{fig_energies} and this feature will be different in the following case where the loops bounce against each other and thus do not merge. The subsequent evolution is then that of a single concentration of magnetic flux evolving as a whole to the top of the domain and keeping its coherence, reaching the surface at around $t=8$ days, as studied in \cite{Jouve3}.

\subsubsection{Influence of field strength and twist intensity}
\label{sect_merge2}

\begin{figure*}[htbp]
	\centering
	\includegraphics[width=17cm]{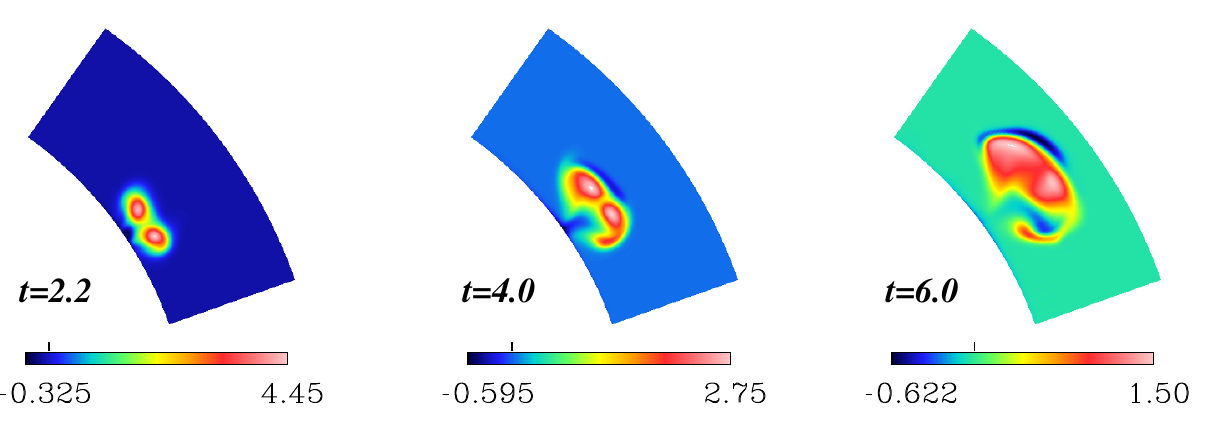}
	\caption{Same as figure \ref{fig_merge} but for CsTsBw. \lau{Because of the weak amplitude of axial field, the attractive Lorentz force is weaker and the merging happens much later than in case CsTsB.}}
	\label{fig_poslamp}
\end{figure*}

\begin{figure*}[htbp]
	\centering
	\includegraphics[width=17cm]{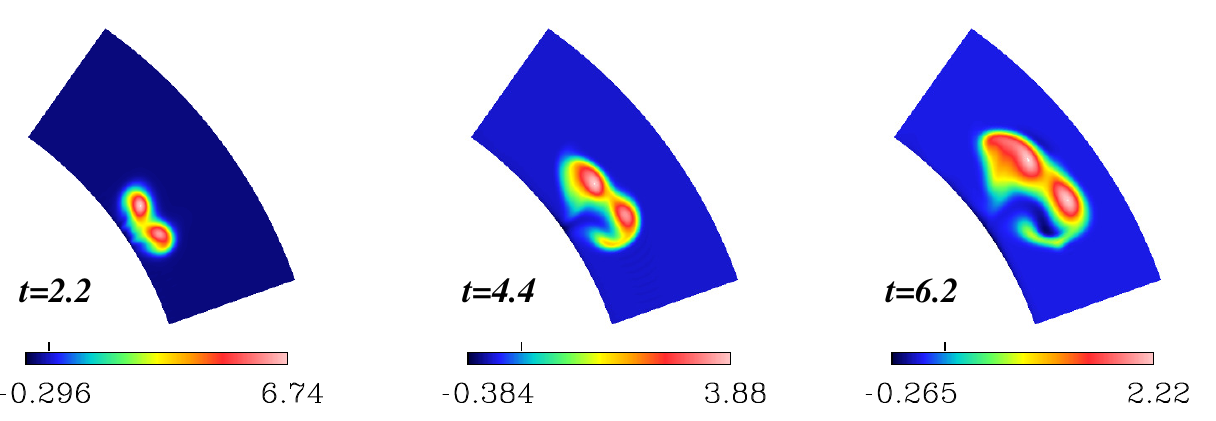}
	\caption{Same as figure \ref{fig_merge} but for CsTwsB. \lau{Because of the weak twist, the Lorentz force is too weak to drag the loops towards each other and the merging does not occur.}}
	\label{fig_posltw}
\end{figure*}

The reconnection between the two loops in the case where the axial field and the twist have the same sign is in fact not generic. It will depend sensitively on the magnetic field strength of the axial component and on the quantity of twist initially introduced. Indeed, as discussed in Section \ref{sect_localc}, the initial latitudinal Lorentz force needs to be sufficiently strong to produce an advection of the loops towards each other before the magnetic diffusion starts to act. Decreasing the initial field strength or modifying the initial twist of the field lines will directly impact the strength of the latitudinal Lorentz force and may produce cases where the loops fail to interact. They would however still be able to interact if they were initially introduced at smaller latitudinal separations. To illustrate this argument, we computed additional cases similar to CsTsB but where the twist of the field lines is divided by two (CsTwsB) and where the amplitude of the axial toroidal field is reduced from $10^5 \rm G$ to $6\times 10^4 \rm G$ (CsTsBw). The structure of the loop sections at about $t=2$, $t=4$ and $t=6$ days are shown on Figures \ref{fig_poslamp} and \ref{fig_posltw} for cases CsTsBw and CsTwsB respectively. In both cases, the merging of the loops which was clearly visible around $t=3$ days in CsTsB shown in Fig.\ref{fig_merge} does not occur. The loops rise next to each other, with only little interaction, especially for the low-twist case (CsTwsB). As we will see later, the loops in CsTsBw do interact but only when they get close to the top of our computational domain.

\begin{figure*}[h!]
	\centering
	\includegraphics[width=8.5cm]{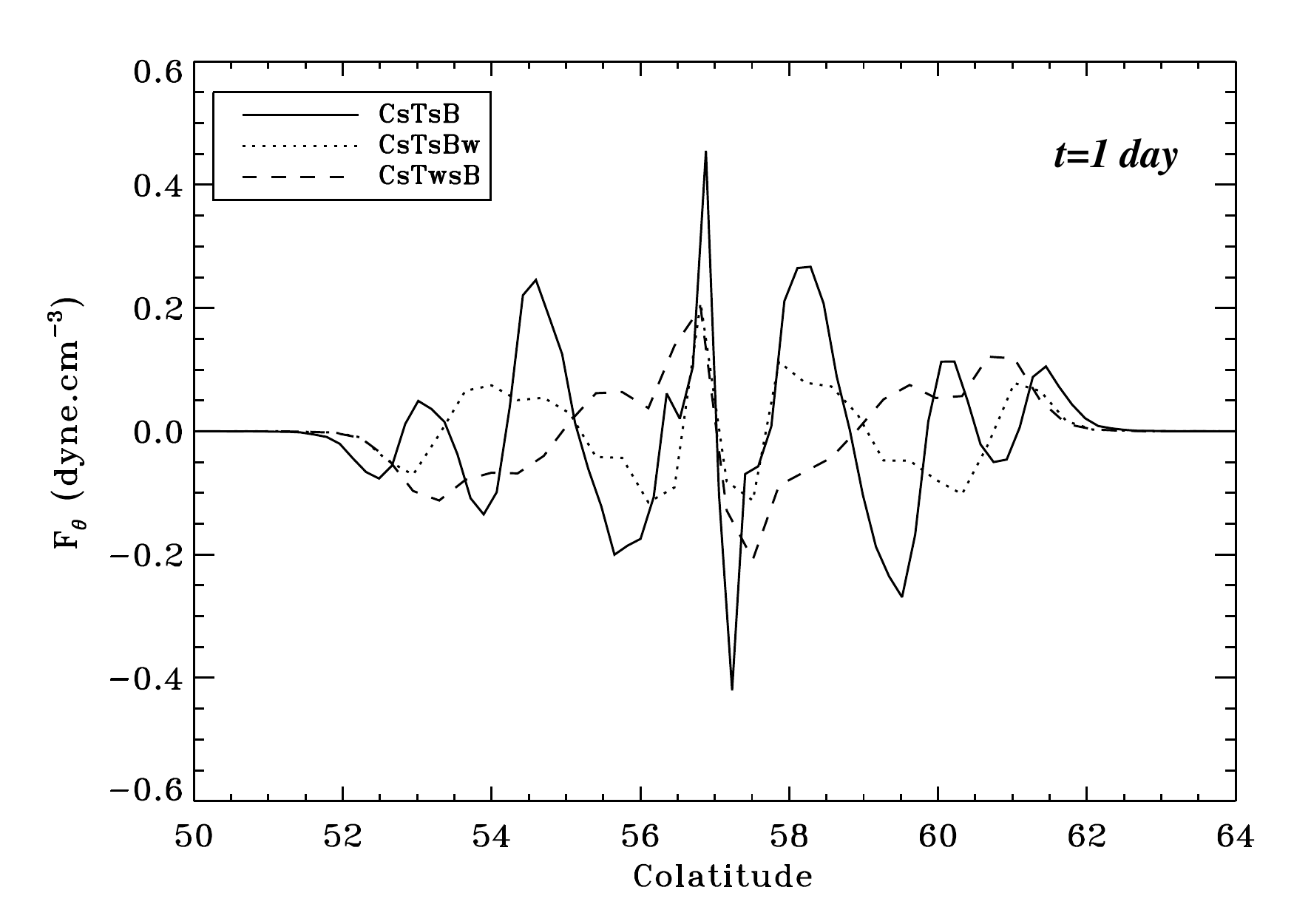}
	\includegraphics[width=8.5cm]{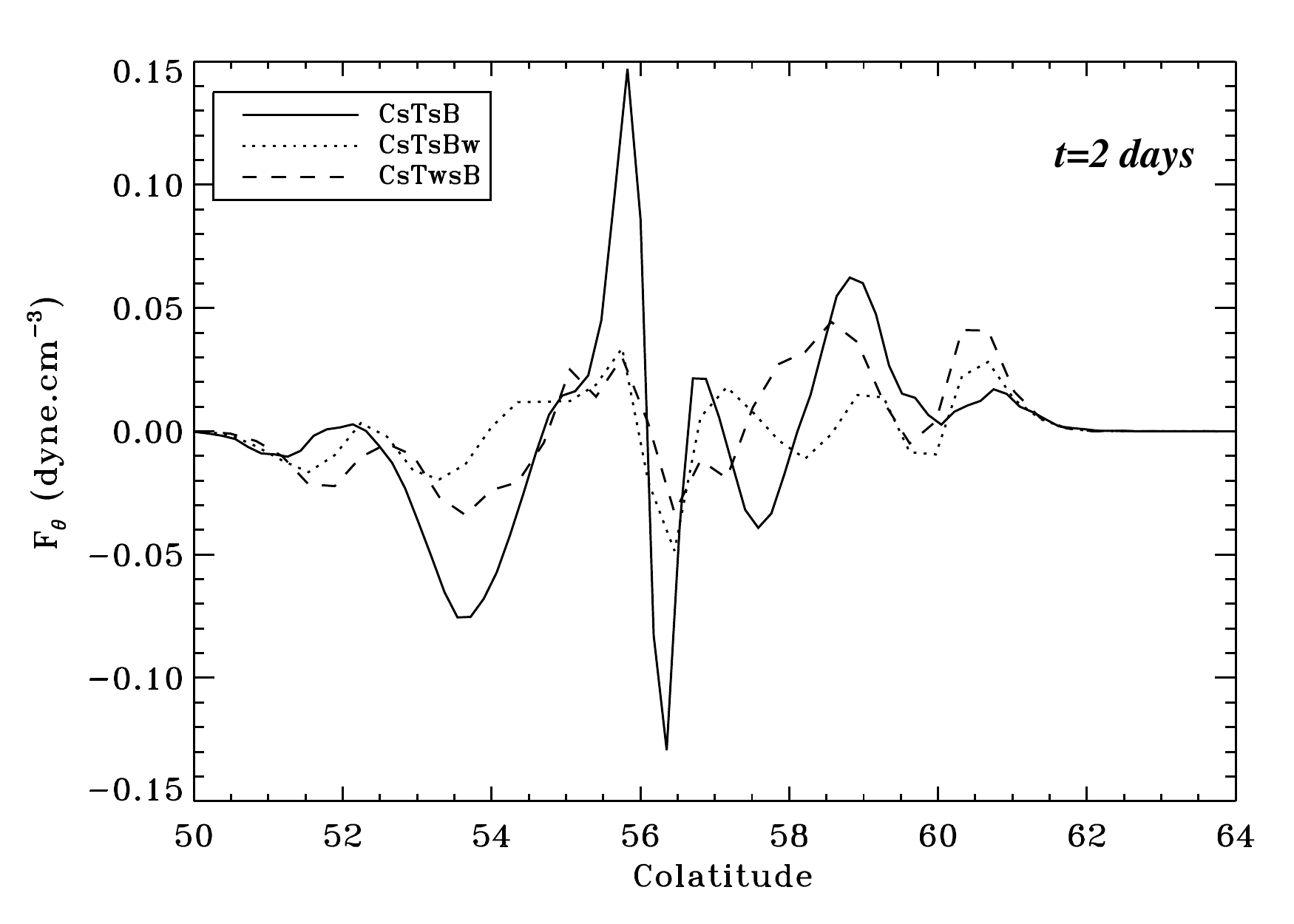}
	\caption{Latitudinal Lorentz force at a longitude of $90^o$ for cases CsTsB, CsTwsB and CsTsBw after 1 day (left panel) and 2 days (right panel) of evolution. \lau{After 2 days, only the case with a sufficiently strong field and twist sustains a strong attractive Lorentz force.}}
	\label{fig_lomerge}
\end{figure*}

This behavior is understandable by looking at the latitudinal Lorentz force which the loop exert around their contact point. 
As we will see in more detail, the reconnection between the loops is enabled by the strong current sheet being created at the periphery of the loops. This strong negative longitudinal current $J_\phi$, created by the strong gradients of the radial magnetic field in the latitudinal direction, gets stronger and stronger as the loops come close to each other. This strong concentration of $J_\phi$, combined with a radial field which changes sign at the contact point between the loops, produces a strong attractive latitudinal Lorentz force which brings the loops closer and closer until they merge in CsTsB. In CsTwsB and CsTsBw, the intensity of the Lorentz force is less because of the reduced axial field or the reduced twist. The loops will thus take more time to get closer to each other. In the meantime, magnetic diffusion will act to reduce both the magnetic field strength and the amplitude of the twist. The loops thus never get close enough to each other and reconnection does not occur. Figure \ref{fig_lomerge} shows the latitudinal Lorentz force at the radius where it is strongest as a function of the colatitude for the three different cases. We can note that after 1 day of evolution, an attractive Lorentz force does exist in the 3 cases. As expected, it is stronger in CsTsB than in CsTwsB and CsTsBw. However after 2 days, it is greatly increased compared to the 2 non-merging cases. \lau{We thus observe two effects here: in CsTsB, the initial Lorentz force dragged the loops closer to each other and thus built a stronger and stronger latitudinal gradient of radial magnetic field, thus producing a large longitudinal current and as a consequence a larger Lorentz force: reconnection eventually occurs. In CsTwsB and CsTsBw, the initial Lorentz force was weaker, the loops were not dragged towards each other sufficiently fast and both the field strength and the twist amplitude have decayed because of magnetic diffusion, the non-linear mechanism of Lorentz force enhancement seen in CsTsB is much less efficient: the loops never get close enough to reconnect in CsTwsB and reconnect much later in CsTsBw.}

\subsection{Opposite twist or opposite axial field: repulsion}

We now investigate two other cases where the field orientation in one loop is modified compared to the previous case. In CoTsB, the twist of Loop 2 is changed from positive to negative, keeping a positive twist for Loop 1. In CsToB, the handedness is kept the same for both loops but the direction of the axial field lines is inverted. In this last case, the two loops thus possess toroidal fields of opposite signs. This is less likely to happen in the Sun than the previous cases since the toroidal field created by the \lau{differential rotation shearing the poloidal field at the base of the convection zone (i.e. the $\Omega$-effect)} is thought to be of constant sign in each hemisphere. However, if the mean poloidal field at the base of the convection zone changes sign within an hemisphere, the $\Omega$-effect acting on it would produce also toroidal fields of opposite signs within this hemisphere. \lau{This could be the case during solar maxima when the poloidal field becomes mostly quadrupolar, at the expense of the dipolar component \citep{derosa12}.} Observations of flux emergence at the Sun's surface also show the common existence of anti-Hale regions where the sign of the leading polarity can be opposite to what is expected during a particular cycle. These anti-Hale regions could originate from flux tubes with an axial field of opposite sign to the loops producing the classical active regions. Finally, if kink instabilities of newly formed buoyant loops occur, twist may be converted to writhe and the axis of the kink-unstable loops may completely change directions (e.g. as simulated by \cite{Torok}).

 In these cases, as we discussed above, the field lines at the loop peripheries will thus be parallel and are not likely to reconnect. Moreover, as we saw in section \ref{sect_localc}, these loops could exert a repulsive force on each other. We want to assess if this analytical prediction is true in full 3D convective simulations.

\subsubsection{Typical case of bouncing loops}

\begin{figure*}[h!]
	\centering
	\includegraphics[width=17cm]{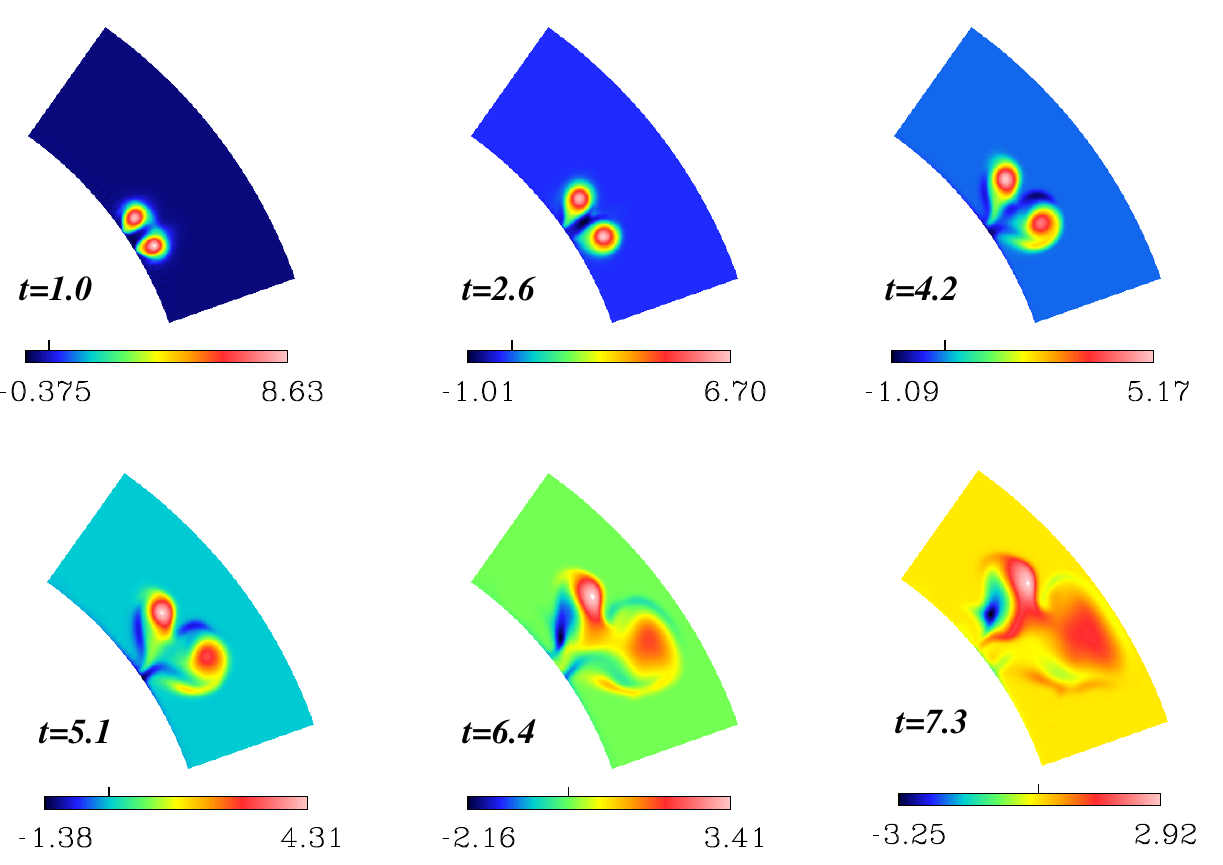}
	\caption{Same as Fig.\ref{fig_merge} but for CoTsB. \lau{Instead of merging, the loops now tend to be pushed away from each other. Note that the field amplitude is also maintained at higher values compared to the merging case of  Fig.\ref{fig_merge}.}}
	\label{fig_bounce}
\end{figure*}

Figure \ref{fig_bounce} shows the typical evolution of two magnetic loops in CoTsB. In this case, the sign of the twist of the field lines in Loop 2 (at higher latitude) is reversed. We note that a similar situation is found for two loops of opposite axial field but with twist parameters of the same sign (CsToB and CsToBw, not shown).

 If we look more closely at the temporal evolution of the loops in the meridian plane in Fig. \ref{fig_bounce}, we find that not only do the loops avoid any reconnection but they tend to repel each other while rising through the convection zone, as predicted by our analytical calculation of the Lorentz force. As a consequence, it is visible at times $t=4.17$ and $t=5.10$ days that the loop initially located at a higher latitude adopts a trajectory almost parallel to the rotation axis when the loop at lower latitude is slightly pushed towards the equator. The advection of Loop 2 towards higher latitudes is enhanced by the poleward background convective velocity field dominating at that longitude and time, as we shall see in Fig.\ref{fig_vtheta}. \lau{Other forces such as hoop stresses also provide a global poleward motion to the loops}. As a comparison to the previous case, we also note that since reconnection did not occur and the magnetic energy intensity did not change because of any interaction between the loops, the magnetic field intensity in the loop by time $t=4.17$ is still of about $52\%$ of the initial magnetic field strength, to be compared to the $37\%$ remaining in the previous CsTsB. This still means that magnetic diffusion is not negligible in these various cases. 

\subsubsection{Comparison with the attractive case}

\begin{figure*}[h!]
	\centering
	\includegraphics[width=17cm]{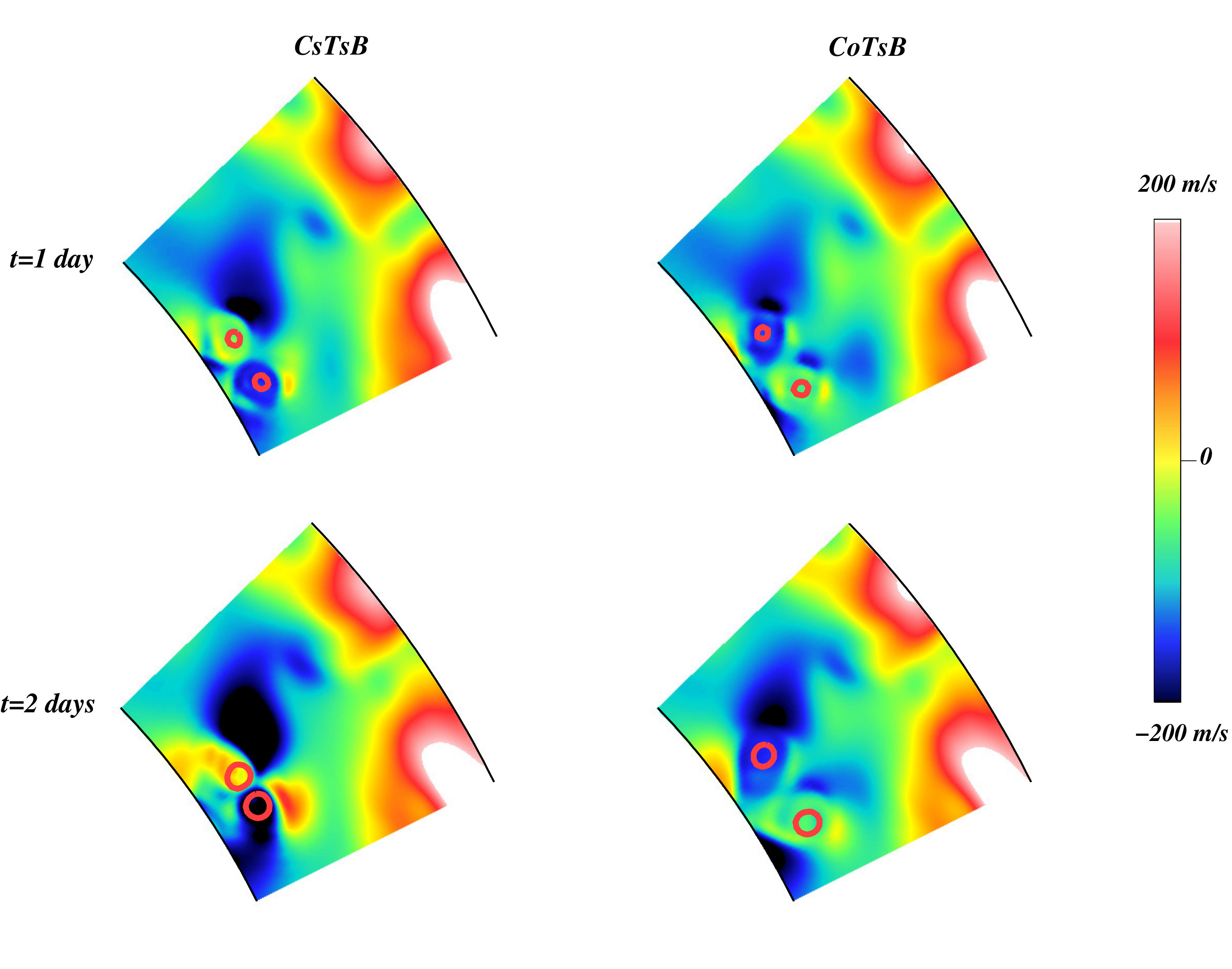}
		\caption{Latitudinal component of the velocity field $v_\theta$ at times $t=1$ (top panels) and $2$ (bottom) days in CsTsB (left panels) and CoTsB (right), at longitude $90^o$, \lau{zoomed in a latitudinal wedge where the loops are located. The red circles indicate contours of $8\times 10^4 \rm G$ toroidal fields on the top panels and of $6\times 10^4 \rm G$ fields on the bottom panels. Note that $v_\theta$ has opposite signs at the loop locations in CsTsB and CoTsB, explaining the merging in CsTsB and bouncing in CoTsB.}}
	\label{fig_vtheta}
\end{figure*}

We analyze in detail in this subsection the origins of the different behavior between this CoTsB and the previous CsTsB. To do so, we plot in Figure \ref{fig_vtheta} the latitudinal velocity in the region where the loops are located, after 1 and 2 days of evolution, in CsTsB (same twist)  and in CoTsB (opposite twist). We note that our loops are clearly embedded in a convective environment and that the velocity field they create through the Lorentz force is lower or of the same order as the background convective flow. The loops are thus affected both by the convective motions and by the velocity field they produce. If we focus on the snapshots after 1 day of evolution, we recover the results we expected from the calculations of the Lorentz force: inside the red circle which marks the presence of the magnetic structures, the flow is such that the loops are attracted towards each other in CsTsB and are advected away from each other in CoTsB. However, the loops are here initially introduced in a globally poleward flow, resulting in all cases in magnetic structures being advected towards high latitudes. We note for example that in CoTsB, the poleward flow created by the Lorentz force at higher latitudes adds to the background poleward velocity field of the convection and Loop 2 (at higher latitude) will thus be strongly advected towards the pole. Loop 1, on the contrary, is pushed away from Loop 2 and should then move equatorward. However, since it is also embedded in a poleward flow slightly faster than the magnetically-induced flow, it is also slightly moving poleward.

At time $t=2$ days, the velocity field due to the Lorentz force becomes stronger, particularly in CsTsB where magnetic reconnection is about to start. Here again in the red circle, it is rather clear that the Lorentz force is strongly attractive between the loops at the radius of their apices. Higher up at the loop peripheries, an opposite flow is created but which only acts on the top parts of the loops, their axis being still strongly advected towards each other. In CoTsB on the contrary, the loops are still pushed further away, but with a smaller velocity field amplitude. The Lorentz force acting between the loops becomes less intense as the loops move away from each other.

\begin{figure*}[h!]
	\centering
	\includegraphics[width=8.5cm]{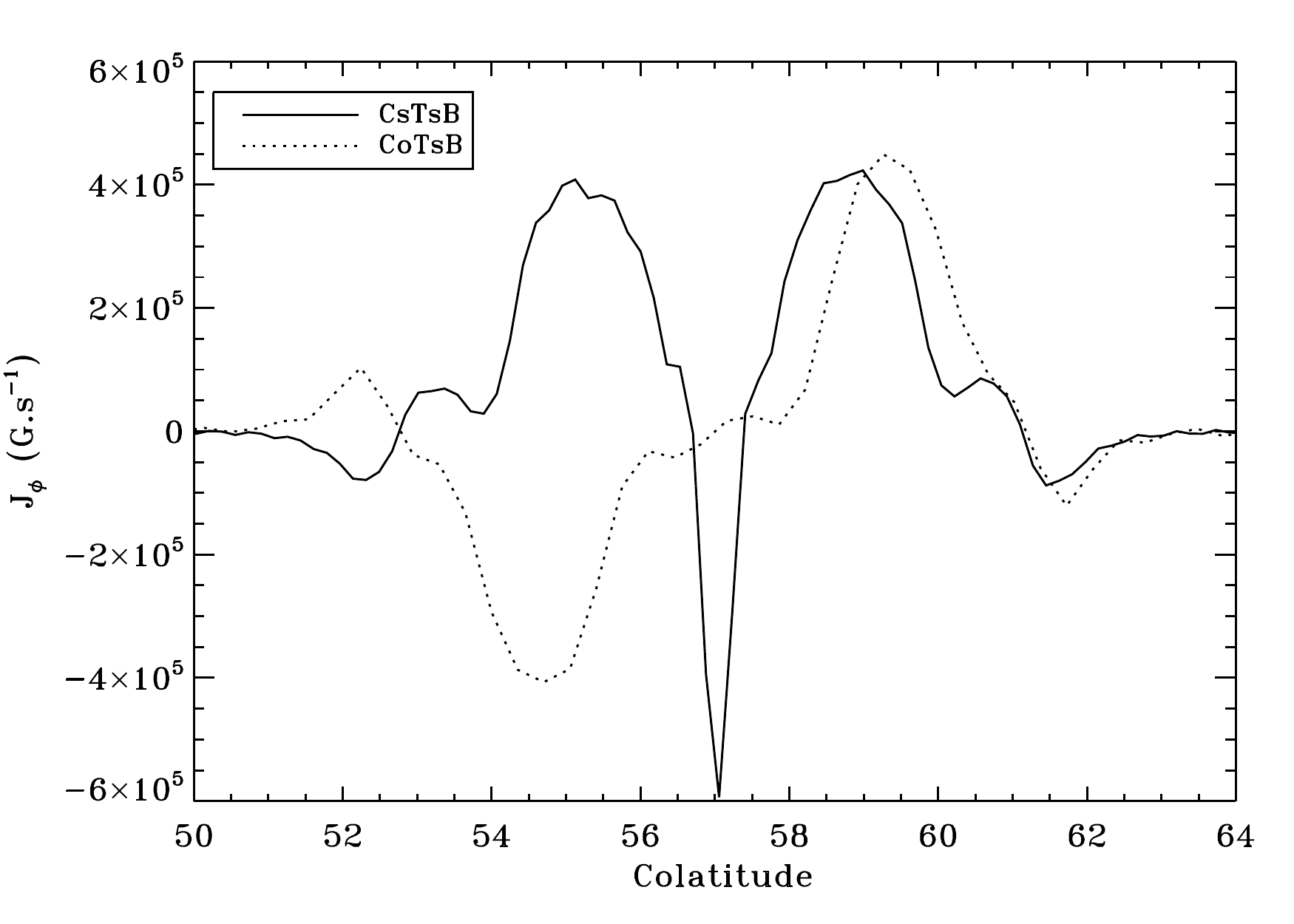}
	\includegraphics[width=8.5cm]{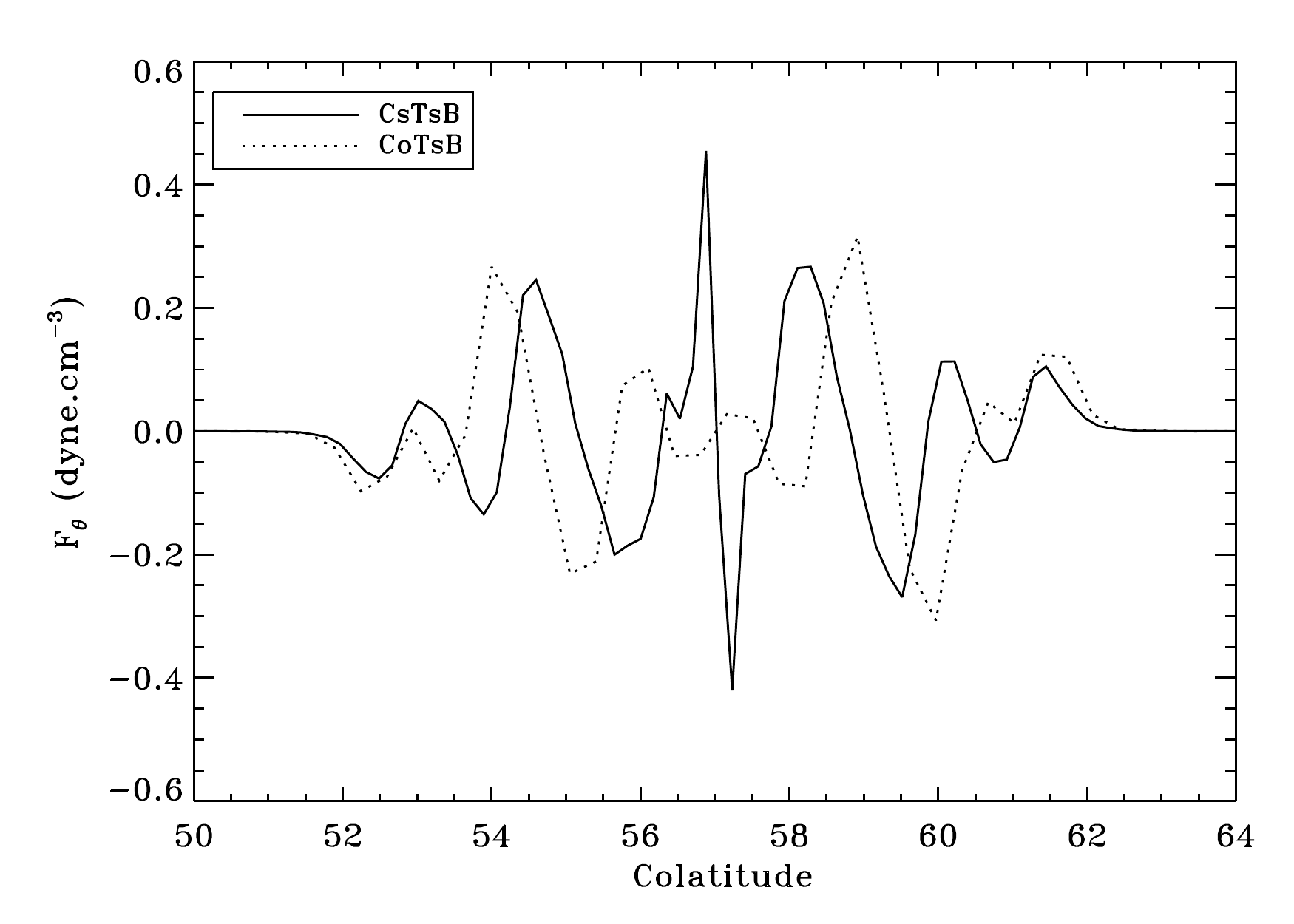}
	\caption{Longitudinal current (left panel) and latitudinal Lorentz force (right panel) after 1 day of evolution of the loops of CsTsB and CoTsB.\lau{In CsTsB, a strong current sheet if formed at the contact point between the loops, enabling them to merge, which is not the case for CoTsB.}}
	\label{fig_lotw}
\end{figure*}

Again in these two cases, it is interesting to plot the Lorentz force since it is responsible for the velocity field observed in the previous figure. 
Figure \ref{fig_lotw} shows a comparison between CsTsB and CoTsB for these quantities of interest. Plotted are the longitudinal current and latitudinal Lorentz force at the radius where these quantities are strongest, as a function of latitude and at the longitude of $90^o$, after 1 day of evolution. If we focus on the longitudinal current, we first note that after 1 day of evolution, the loops in CsTsB have already moved much closer together than in CoTsB. Indeed the two positive maxima of $J_\phi$ in CsTsB corresponds to the location of the loop axis and the strong negative current is located at the contact point between the loops. After 1 day of evolution, the loops are thus located at colatitudes of around $56.5^o$ and $58.5^o$ (we recall that the initial colatitudes were $56^o$ and $60^o$). For CoTsB, the loops have already started to repel each other, being after 1 day located at the colatitudes of $54.5^o$ and $59.5^o$. In particular, it is clear that the loop located at higher latitude has moved even higher up. We note again here that all the loops move to higher latitudes, even in the repulsive cases where the loops are pushed away from each other. This is due to the background poleward convective flow in which the loops are embedded. Moreover, the strong negative longitudinal current which was visible for CsTsB does not exist for CoTsB. The main term responsible for the generation of $J_\phi$ here is $-1/r \,\, \partial_\theta \,B_r$. In CsTsB, the radial field increases as a function of colatitude at each side of the contact point between the loops, thus creating a globally negative $J_\phi$. On the contrary, in CoTsB, the radial field decreases at the periphery of the higher loop and then increases again at the periphery of the other loop, thus creating longitudinal currents of opposite sign which quickly cancel each other. We are then left with a very small $J_\phi$ at the contact point between the loops.

The latitudinal Lorentz force is plotted on the right panel of Fig.\ref{fig_lotw}. As expected, a strong latitudinal Lorentz force is created at each side of the contact point between the loops, positive at higher latitudes and negative at lower latitudes, thus implying an attractive force between the magnetic structures. On the contrary, in CoTsB, the radial field keeps a positive sign across the boundary. The longitudinal current is slightly negative at higher latitudes and slightly positive lower, thus producing a negative Lorentz force at high latitudes and a positive one lower. This produces a weak repulsive force between the loops, explaining the tendency for the magnetic structures to be repelled from each other while they rise.

\subsection{Opposite axial field and opposite twist: full reconnection?}

We now investigate the last case, where both the axial field and the twist of the field lines are of opposite signs in the two loops. This corresponds to CoToB and CoToBw described in Table \ref{tab}. In these cases, a reconnection is expected to occur first between the components of the magnetic field corresponding to the twist and then between the axial fields which point in opposite directions. 

\begin{figure*}[h!]
	\centering
	\includegraphics[width=17cm]{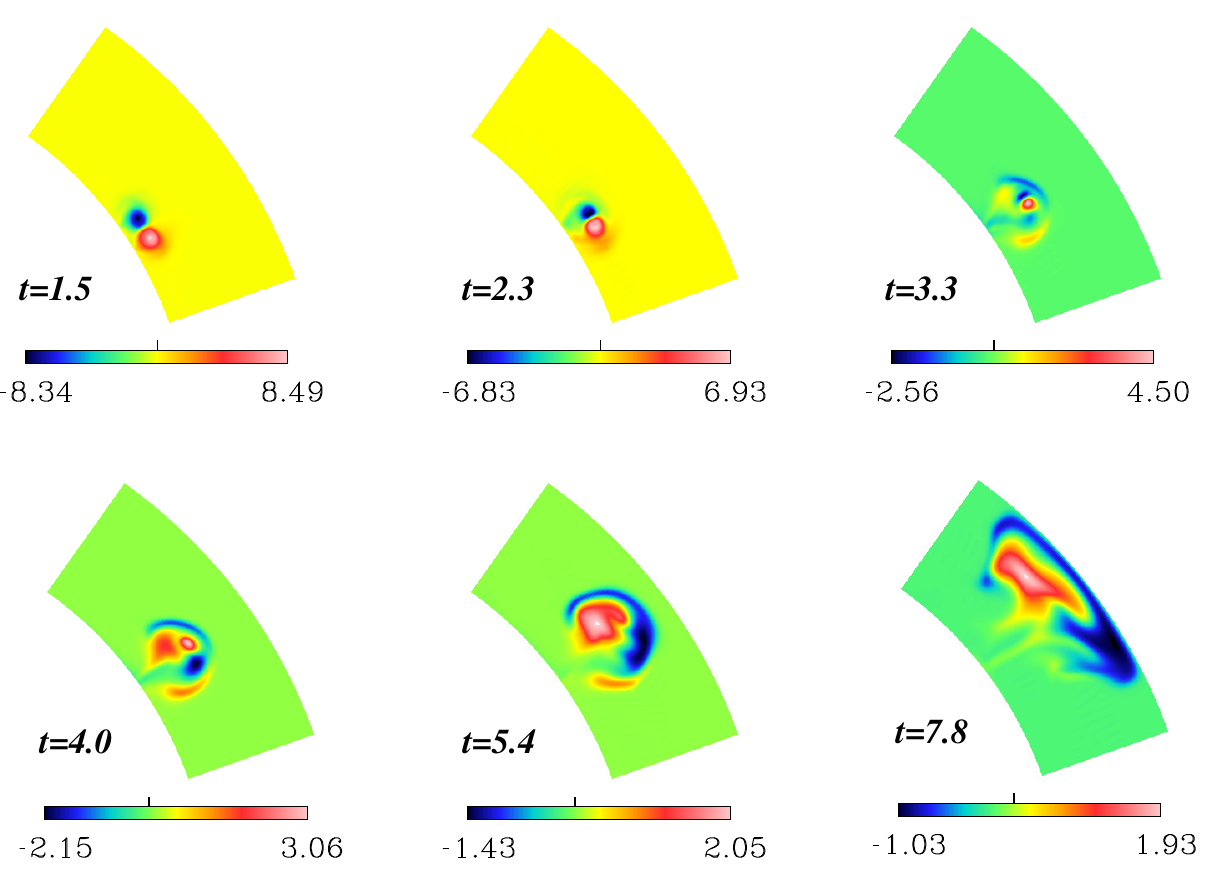}
	\caption{Same as Fig.\ref{fig_merge} but for CoToBw. \lau{In this case, a strong interaction occurs between the structures: the loop of positive $B_\phi$ passes through the loop of negative $B_\phi$ after 3 to 4 days of evolution.}}
	\label{fig_recon}
\end{figure*}

Figure \ref{fig_recon} shows the temporal evolution of such loops. The evolution is indeed significantly different from the previous cases. During the first 2 days of evolution, the loops get closer together, due to an attractive Lorentz force in the same way as in CsTsB above. The loops then merge and seem to cross each other, since the positive toroidal field of the loop located at lower latitudes ends up appearing at higher latitudes at $t=4$ days and later. From the temporal evolution of the energies which were shown in Figure \ref{fig_energies}, it was clear that this case was the one producing the largest release of magnetic energy because of the multiple reconnections and thus the largest increase in kinetic energy. However, the loops do not seem to fully reconnect since the two main initial structures are still visible after 8 days of evolution. What is however clear is that the loops underwent a strong loss of magnetic intensity. Indeed, after 4 days, only $30\%$ of the initial magnetic intensity is left in the loops and only $20\%$ after $5.4$ days. As we shall see in the next section, even if the magnetic structures still seem to be well confined in latitude and radius, the morphology of the emerging radial field is very different from the other cases.

\section{Consequences for emerging regions}

One of the main motivations to study interacting $\Omega$-loops was to investigate the consequences of such interactions on emerging magnetic regions at the solar surface. We note here that in our calculations, the top of the domain is located at $r=0.96 R_\odot$, i.e. still quite far from the solar surface but we can get some insight on the possible implication for solar active regions by looking at the radial magnetic field close to the top of our domain in each case. \lau{We first focus in this section on the top panels of Figs. \ref{fig_shslbounce}, \ref{fig_shslmerge}, \ref{fig_shslmergew}, \ref{fig_shslmergeisen} and \ref{fig_shslfull} and will discuss the bottom panels (mainly the radial current maps) in the next section.}

\subsection{Bouncing loops}

We first examine the radial magnetic field produced by the emergence of loops which do not merge but instead repel each other while rising. Since this situation is not really likely to produce a unique active region with an intricate pattern of positive and negative polarities but instead two separate bipolar magnetic regions, we will not particularly focus on this case. However, it is interesting to see what kind of structure is obtained at the top of the domain, to be then compared to the merging cases presented in the next section.

\begin{figure*}[h!]
	\centering
	\includegraphics[width=17cm]{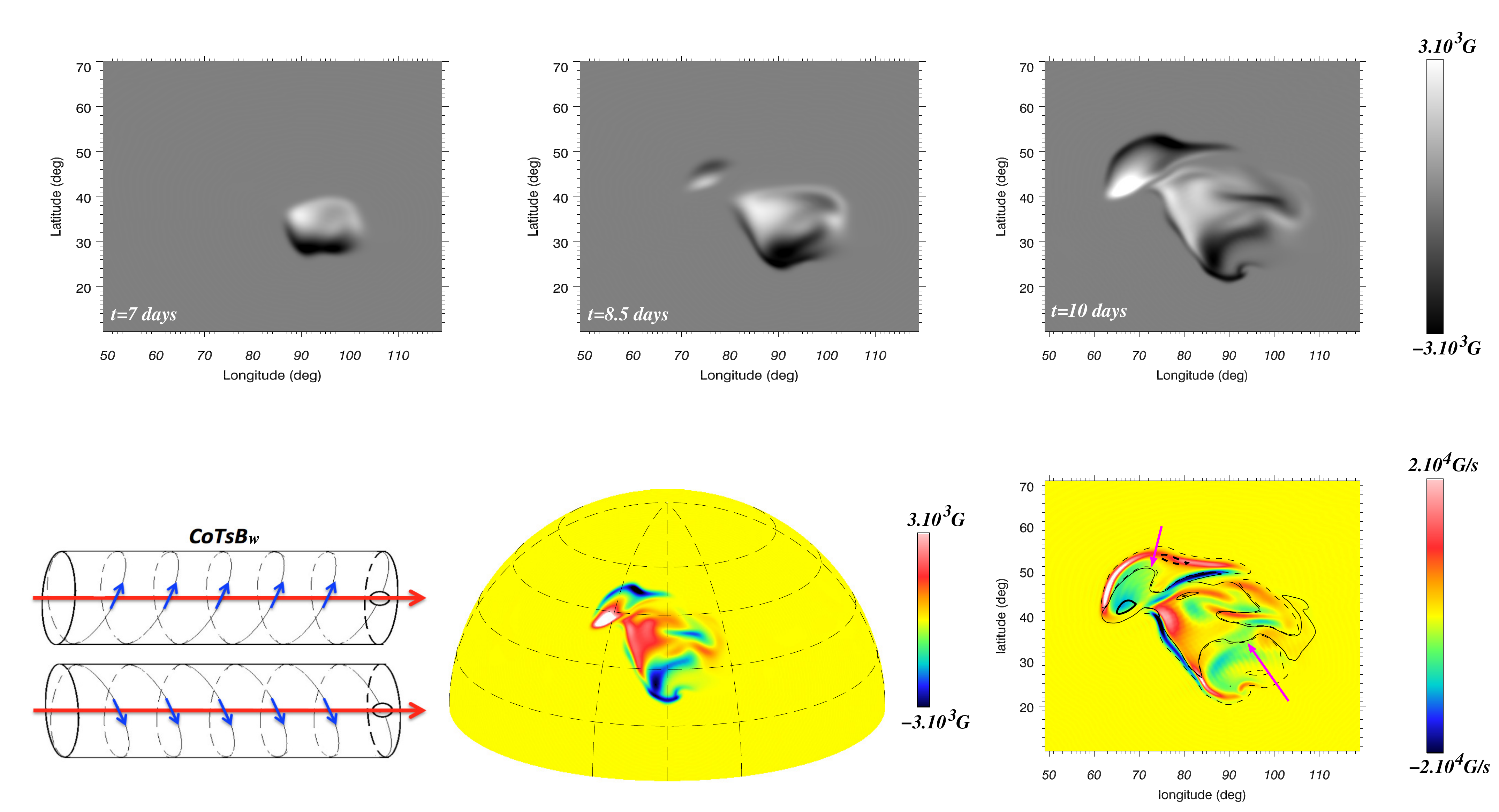}
	\caption{Case CoTsBw: top panel: zoom on the emerging radial magnetic field at $r=0.93 R_\odot$ at three different times. Bottom panel: sketch of the initial configuration, emerging radial field at $r=0.93R_\odot$ shown on the whole Northern hemisphere and zoomed map of the radial current $J_r$ at the last time $t=10$ days. On this last panel, we superimpose contours of the radial magnetic field at $80\%$ (thicker) and $2\%$ (thinner) of its min and max values. The PIL is then visible when the contours at $2\%$ of opposite sign meet and is indicated by the magenta arrows. \lau{Two emerging regions of different handedness are here clearly visible.}}
	\label{fig_shslbounce}
\end{figure*}

The 'bouncing loop' situation is obtained in the cases where the axial fields have the same direction but the twists are opposite (CoTsB and CoTsBw) and where the axial fields are opposite but the twists are identical in both loops (CsToB and CsToBw). The results are shown in Figure \ref{fig_shslbounce}. The first snapshot at around $t=7$ days shows the appearance at $r=0.93R_\odot$ of the radial field coming only from the first loop. This first loop to emerge is Loop 1 which was slightly more buoyant than Loop 2 according to the initial entropy perturbation ($A_{S1}=200$ and $A_{S2}=150$). At time $t=8.5$ days (mid-panel), we now see both emerging regions, which are clearly spatially separated. We note that the second loop emerges at rather high latitudes (the polarity inversion line being located at around $45^o$ of latitude) compared to the latitude of introduction (namely $34^o$). The fact that Loop 2 was pushed towards higher latitudes was already observed in Figure \ref{fig_bounce}. On the contrary, Loop 1 emerges almost as if it was introduced alone. Finally, we note that around $t=10$ days, the tilt angle of the high-latitude bipolar structure in CoTsB is opposite to the bipolar region produced by Loop 1 (and is also opposite to the tilt angles found in CsToB, not shown). \lau{This is explained by the fact that this tilt is due to the direction of twist of the field lines. The twist of the field lines will naturally produce a tilt in the structure of the emerging radial field (see \cite{2008ApJ...676..680F} and \cite{Jouve3} for more details about the twist-induced tilt).} In CoTsB, the twist of Loop 2 is left-handed and the axial field is positive, so that the positive radial field is located in the trailing spot, at lower latitudes than the leading negative polarity. On the contrary in CsToB, the twist is right-handed thus producing a tilt angle of the same sign as the emerging region produced by Loop 1 but since the axial field is negative, the leading polarity is positive while the trailing polarity is negative.

\subsection{Merging loops}

We now focus on the more interesting cases in terms of complexity of emerging magnetic regions, i.e. the merging cases. These correspond to cases CsTsB and CoToB where reconnection is allowed because of the presence of anti-parallel field lines.

\subsubsection{CsTsB: same handedness, same direction of axial field}
\label{sect_csts}

\begin{figure*}[h!]
	\centering
	\includegraphics[width=17cm]{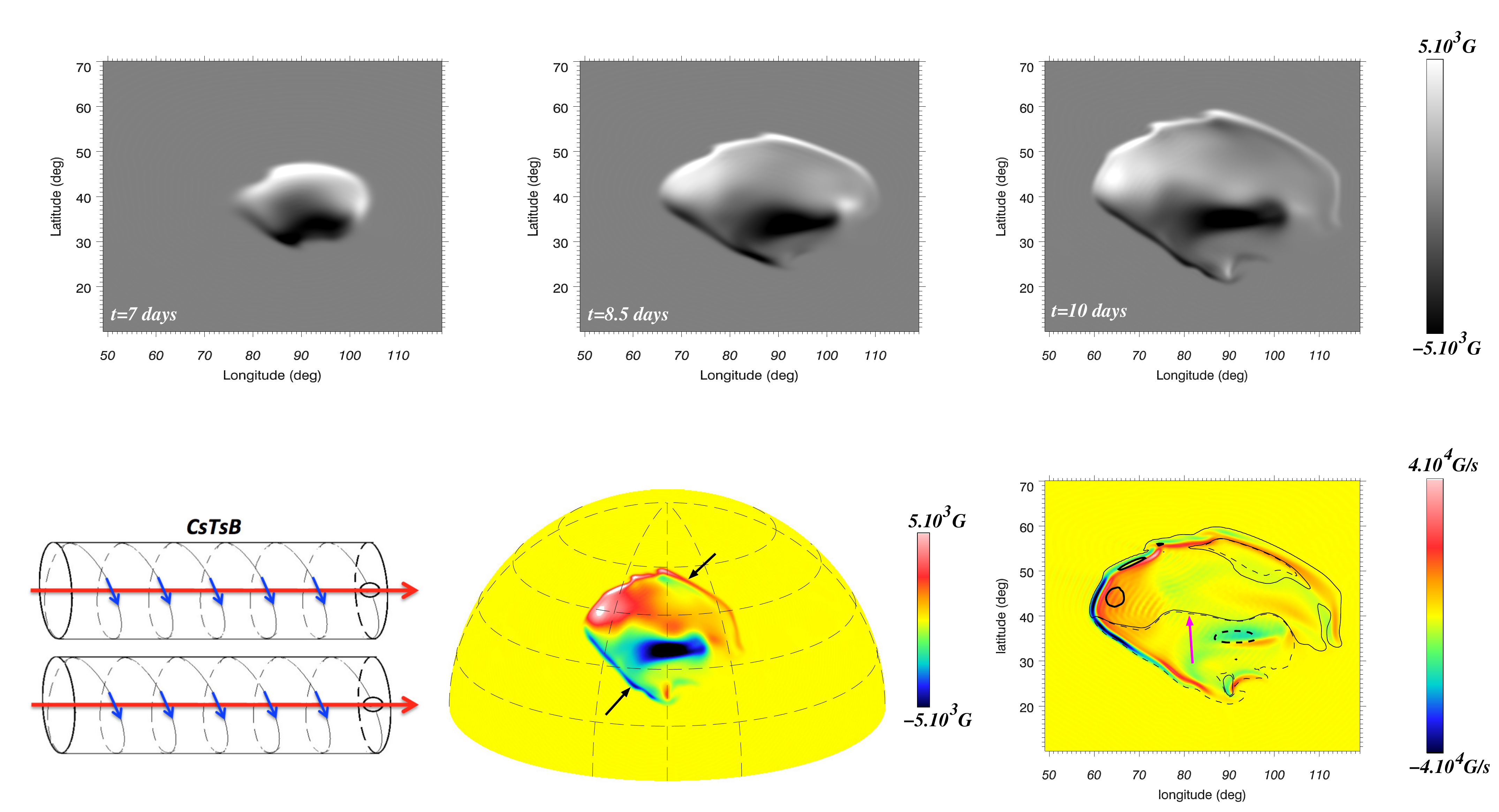}
		\caption{Same as Fig. \ref{fig_shslbounce} but for case CsTsB. \lau{The black arrows on the bottom mid-panel indicate the locations of strong magnetic ``necklaces". In this merging case, the emerging radial field mostly consists of one large bipolar structure.}}
	\label{fig_shslmerge}
\end{figure*}

\begin{figure*}[h!]
	\centering
	\includegraphics[width=17cm]{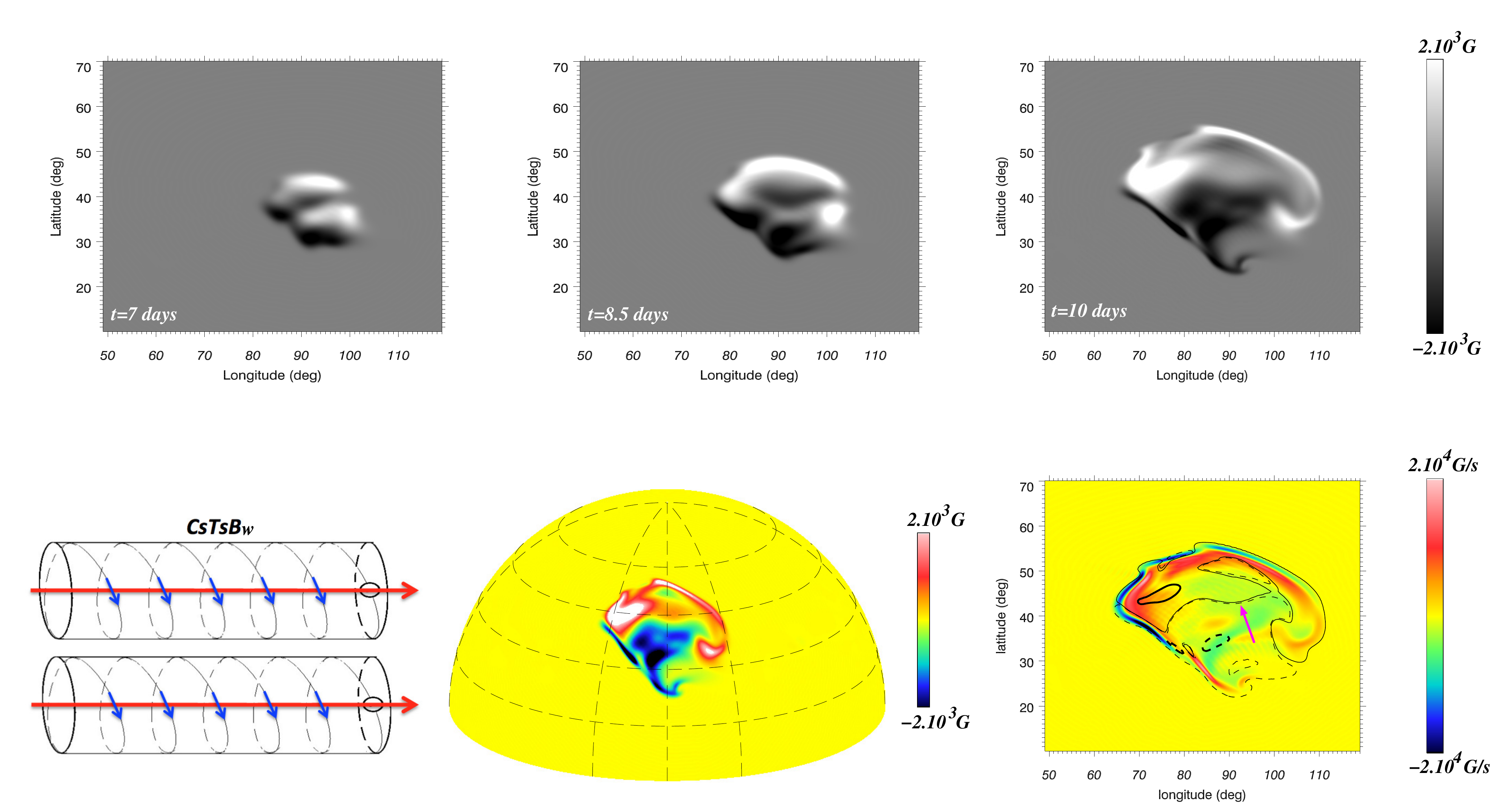}
	\caption{Same as Fig. \ref{fig_shslbounce} but for case CsTsBw. \lau{In this weaker field case, the magnetic structure has been more affected by convective motions and the PIL thus appears more sheared and fragmented than the previous case}.}
	\label{fig_shslmergew}
\end{figure*}

We first consider the cases where the two loops have the same direction for the magnetic field lines, i.e. same twist and same axial field. This corresponds to CsTsB and CsTsBw. The difference between those cases is the initial magnetic field strength, $B_0=100 \rm kG$ for CsTsB and $60 \rm kG$ for CsTsBw. As we saw in the preceding section, the initial field strength will produce an attractive Lorentz force of various intensity which can make the loops merge more or less quickly. In CsTsBw, the attractive Lorentz force is weak enough so that the loops hardly merge before they reach the top of our computational domain, as seen in Sect. \ref{sect_merge2}. Figures \ref{fig_shslmerge} and \ref{fig_shslmergew} show the radial field emerging at $r=0.93 R_\odot$ for CsTsB and CsTsBw. 

At the early stages of emergence, the difference between CsTsB and CsTsBw is rather obvious. In the first case, the merging between the loops has occurred at around $t=3$ days whereas in the second case, the loops emerge in two separate regions since the full merging between the loops has not happened yet. In the top \lau{left panel of Fig.\ref{fig_shslmergew}}, we thus observe two separate emerging bipolar structures, located between the latitudes of $30^o$ and $45^o$, i.e. quite close to the latitudes of introduction. On the contrary, on the top left panel of Fig.\ref{fig_shslmerge}, an individual emerging region appears, with a tilt angle of almost $90^o$ due to the strong right-handed twist initially introduced in both loops and which is maintained in the merged structure. However, it is noticeable that the twist-induced tilt is lower in this case than in the single loop case since reconnection between the loops may have produced dissipation at small scales. 

As the emergence progresses, the two polarities tend to diffuse away from each other in CsTsB and to create a finer scale structure at the periphery of the main concentrations of radial magnetic field (\lau{shown by the black arrows on the bottom mid-panel of Fig.\ref{fig_shslmerge}}). These elongated small-scale structures are also reminiscent of what was observed in simulations of the rise of individual loops studied in \cite{Jouve3}. Indeed, those regions, which were called "magnetic necklaces" were identified as regions of strong vorticity and thus strong shear. We note again, as in \cite{Jouve3}, that the necklaces appear only after a significant fraction of the loops have emerged, since the strong vorticity is located at approximately the same radius of the loop axis. In CsTsBw, the two separate loops continue their emergence in relatively distinct areas but do show signs of merging between identical polarities.

At time $t=8.5$ days, the two separate regions are indeed hardly distinguishable and at the end of the simulation around $t=10$ days, the difference with CsTsB is much less obvious. The elongated small-scale regions of strong vorticity are again observed and two main concentrations of radial magnetic field of opposite polarities now dominate. We note however that the negative polarity is more diffuse in CsTsBw than in CsTsB. 

Another difference between those 2 cases is the extension of the active region. Indeed, at 10 days, the magnetic field occupies about $60^o$ in longitude and $40^o$ in latitude in case CsTsB, whereas the extension is of $40^o$ in longitude and $30^o$ in latitude in case CsTsBw. This is due to the fact that since the magnetic field is stronger in CsTsB, the back reaction on the flow is also stronger. The extension of the convective cells while the loops reach $r=0.93R_\odot$ is thus stronger in CsTsB, explaining the larger extent of the active region. 
\lau{The idea of this work was to perform a large number of simulations to investigate the various scenarii for flux tube reconnection, and not to reproduce in detail observed solar active regions. This motivated our choice of relatively large radius and longitudinal extent of our buoyant loops. The areas occupied by our simulated emerging regions thus reach a few $10^4$ millionths of solar hemisphere (MSH), which is quite large compared to observed solar active regions, typically covering around a few $10^2$ MSH. However, as shown in \citet{Aulanier13} (in particular their Fig.2), the areas of the largest active regions can easily reach a few thousands of MSH: a value of about 6000 MSH has been reported for the largest active region ever observed in April 1947 \citep{Nicholson48,Taylor89}. According to the model of \citet{Aulanier13}, the typical magnetic flux of such a region could reach a value of $2\times10^{23} \rm Mx$, i.e. not so different from what we get in our simulations (see section \ref{sect_flux} below). Moreover, since the top of our domain is still about $28 \rm Mm$ away from the photosphere, we could argue that small-scale convective motions higher up can still pin down some portions of the emerging structure and lift some others so that only a part of our very large region will indeed show up a the photospheric level. A connection with compressible simulations of the upper convection zone should be investigated to address this question, as done for example in \citet{2017ApJ...846..149C}. This lies beyond the scope of the present work.}

%\lau{We note that the emerging regions simulated here are very large compared to solar active regions, whose extent never reach more than $10^o$ in latitude and longitude at the photosphere. This is related to the characteristics of our global numerical model in which artificially large flux tubes have to be introduced to survive diffusive effects and to our choice of entropy perturbation extending on $15^o$ in longitude. 
%We also find that the shift towards lower longitudes is more pronounced in CsTsB. Indeed, the loops develop a retrograde flow along their axis to ensure angular momentum conservation and the intensity of this retrograde flow is also increased when the field strength is increased. As a consequence, the retrograde motion of the loops is expected to be stronger in CsTsB than in CsTsBw and the longitude of emergence thus shifted towards lower values, as seen on the top panels of Figs. \ref{fig_shslmerge} and \ref{fig_shslmergew}.

It is interesting to compare the results of CsTsB with the emerging region obtained in a simulation of a single $\Omega$-loop (not shown here). The major differences lie in the extension of the emerging structure in latitude and longitude and in the shape of the main concentrations of field, especially of positive polarity. The extension in longitude is larger in CsTsB than in the single loop simulation. This is to be expected since the loops where introduced with a deficit of entropy on a $15^o$ extension but peaking at two different longitudes, namely $100^o$ and $92^o$. If we moreover take into account the expansion of the magnetic structures during their rise, we indeed end up with an extension in CsTsB of around $35^o$ in CsTsB and around $25^o$ in the single loop case. It is finally interesting to note that the necklaces are much more pronounced in the merging cases here than they were in the single loop simulations and than they are in the non-merging cases, as seen on Fig.\ref{fig_shslbounce}. This indicates that an even stronger vorticity generation is at play in the merging cases, or that the velocity shear acts more efficiently on the peripheral magnetic field which could have been weakened by the reconnection.

\subsubsection{Effects of convection in CsTsB}
\label{sect_cstsbisen}

\begin{figure*}[h!]
	\centering
	\includegraphics[width=17cm]{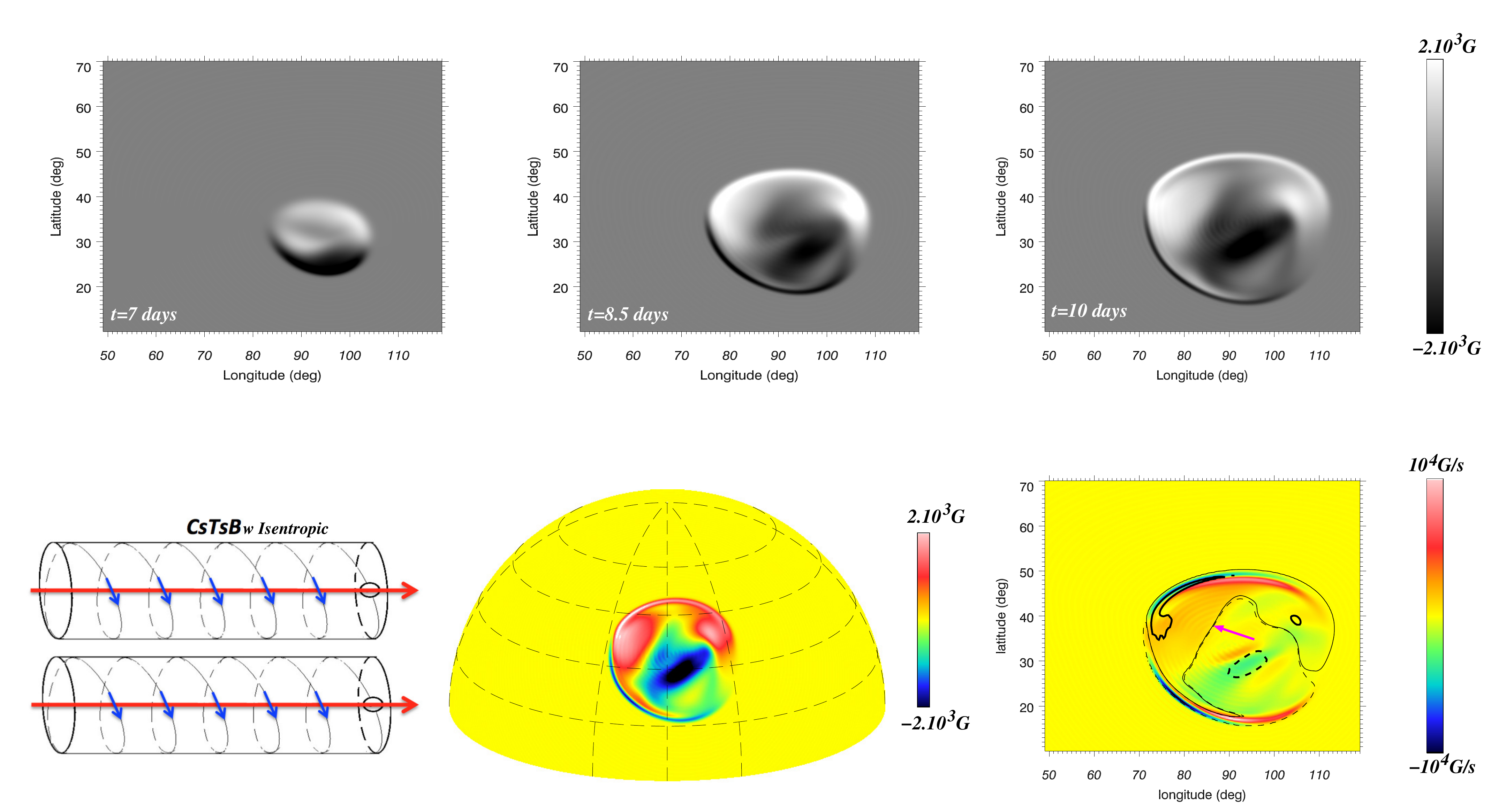}
	\caption{Same as Fig. \ref{fig_shslbounce} but for case CsTsBw isentropic. \lau{Without convection, the emerging structure consists in one large oval-shaped bipolar region, with simple radial field and current structures.}}
	\label{fig_shslmergeisen}
\end{figure*}

In order to investigate in detail the effects of convective motions and associated mean flows like differential rotation and meridional circulation on the rising structures and on the subsequent emerging regions, we computed the same case CsTsBw, without triggering convection. In this case, the spherical shell rotates as a solid body at the imposed solar rotation rate and no significant meridional flow is produced. Since the buoyancy of the loops is mainly controlled by the entropy perturbation initially applied to the flux tubes (and not by the convective up flows), the rise times are fairly similar between the convective and non-convective cases. Moreover, the expected interaction between the loops being related to the orientation of the field lines, they should not be very different between the isentropic and the convective cases. However, the initial attraction between the loops can be tempered by the advection by convective motions. The amplitude of the background convective flows being of the order of that of the velocity produced by the Lorentz forces associated with the loops, we expect a significant advection of the magnetic structures by the surrounding flows. This is indeed what is observed when Fig. \ref{fig_shslmergew} (corresponding to the convective case) is compared to Fig. \ref{fig_shslmergeisen} (corresponding to the isentropic case). 

A close look at the top left panel of both figures allows us to see that the merging between the loops has been more effective in the isentropic case. Indeed, at $t=7$ days, the pattern of the emerging radial field exhibits one large bipolar structure in the isentropic case while two separate regions were still visible in the convective case. This is due to the fact that the merging of the loop is faster in the isentropic case. Indeed, depending on the horizontal divergence or convergence of flows during the rise, the loops may have been pushed apart by convective motions in the first case, while these motions are absent in the second case.

The most striking difference in the radial magnetic field structure is the very circular shape of the emerging region in the isentropic case, compared to the convective one. Indeed, the loops create their own large upflow around their apices while they rise. The edges of this upflow thus naturally follow the contours of the rising magnetic concentration, which is dictated by the initial perturbation applied to the fluxtubes. This perturbation being gaussian in $\theta$ and $\phi$ (with a larger extension in $\phi$ because of the longitudinal separation between the apices), the upflow created by the rising structure takes an oval shape which is here visible on the radial magnetic field. On the contrary in the convective case, this upflow is of the same order as the convective velocity amplitudes and the magnetic structures are thus affected by the advection by those flows. This explains the much less simple shape of the emerging regions of Fig. \ref{fig_shslmergew}. 

A global poleward advection of the whole structure is also visible on the bottom mid-panel of Fig. \ref{fig_shslmergew} when convection is present, which is not the case on Fig. \ref{fig_shslmergeisen}. This is due to the poleward meridional flow sitting in the Northern hemisphere in the convective case during the whole duration of the simulation. Not only meridional flows play a role to advect the magnetic structure here, but also the effect of the differential rotation is visible. In the convective case where the equator rotates faster than the poles, the main positive and negative polarities are shifted towards lower longitudes compared to the rigidly rotating isentropic case. Indeed, at the latitudes of introduction, the rotation is retrograde compared to the rotation of the reference frame. Another major difference is the intensity of the radial field in each polarity which is higher in the convective case, as seen clearly on the bottom mid-panel. This indicates that there is an efficient amplification of the field in the main polarities due to the compression of the field lines by convective motions. Finally, we note that the strong necklaces are still visible in the isentropic case, mostly concentrated again at the periphery of the rising structure, as was the case for CsTsB.

\subsubsection{CoToB: opposite handedness, opposite direction of axial field}

Finally, we present the results on cases CoToB and CoToBw where both the axial field and the twist is opposite in both loops, allowing reconnection of the peripheral magnetic field lines and of the axial field lines. Since the impacts of convective motions are stronger in CoToBw, the results are presented for this weaker field strength case.

\begin{figure*}[h!]
	\centering
	\includegraphics[width=17cm]{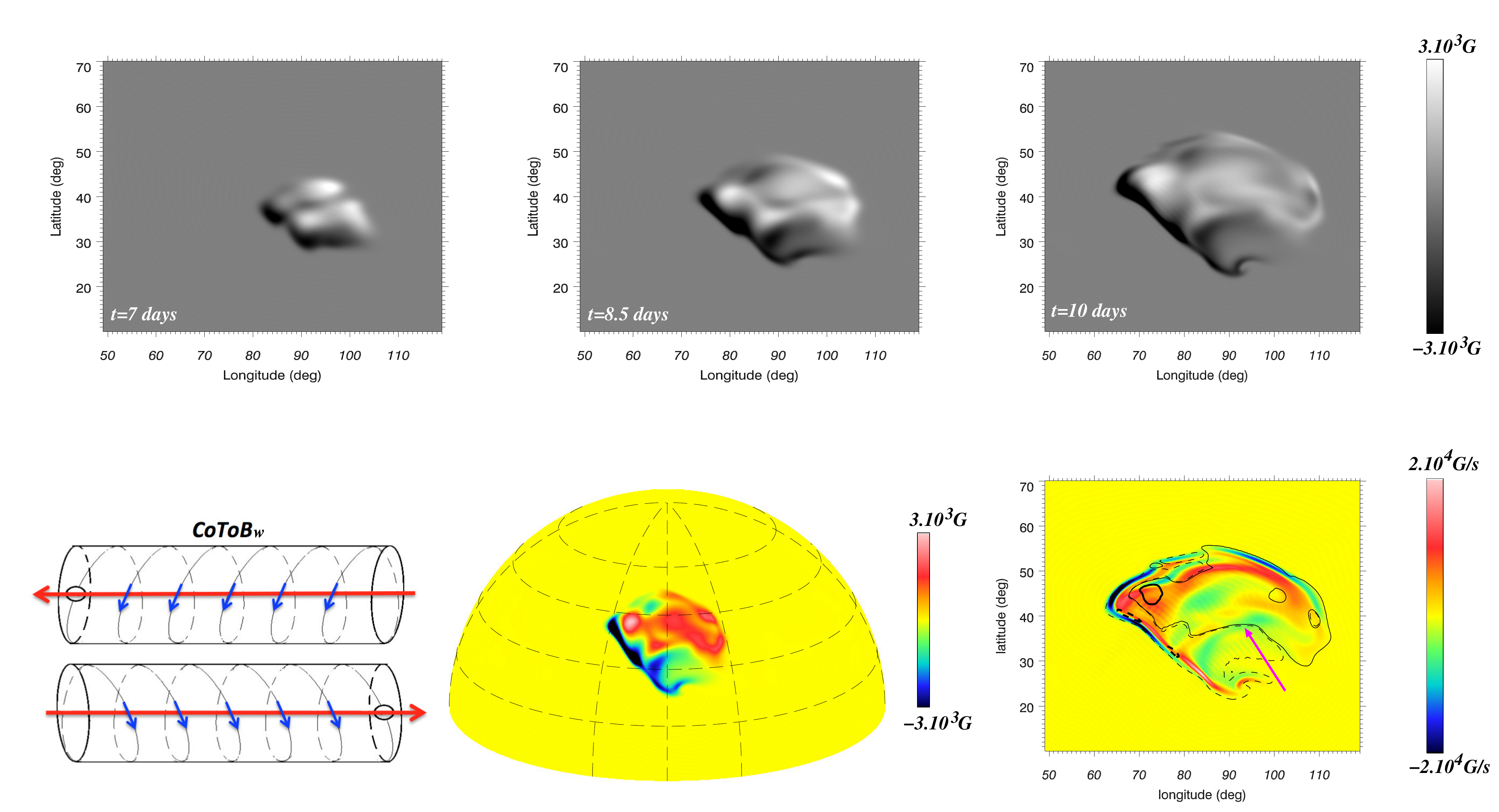}
	\caption{Same as Fig. \ref{fig_shslbounce} but for case CoToBw. \lau{The two polarities are here less concentrated, the negative polarity is particularly elongated and advected by a strong and narrow convective down flow.}}
	\label{fig_shslfull}
\end{figure*}

The results for CoToBw are shown in Figure \ref{fig_shslfull}. The most striking finding here is that the main quasi-circular-shaped concentrations of radial magnetic field in a positive and negative polarity as was found in the previous cases, have partly disappeared. The two main polarities, which were visible in all previous cases, are now much more diffused. We are mostly left only with a weak diffuse radial field sitting on the very large upflow created by the emerging magnetic field itself, and narrow elongated regions of radial field which get advected in the downflow lanes at the periphery of this large positive radial velocity cell. The interaction between the loops in this case thus had the effect of canceling the opposite polarities probably through reconnection between the field lines and this results in the quasi-absence of any bipolar structure which could be associated with a sunspot at the solar surface. It is unclear however if this situation is similar to what \cite{Linton} reported as a "slingshot" reconnection which drags the magnetic structures more horizontally than vertically. However, we do get a similar strong flux annihilation, which is also seen in the faster drop in magnetic energy in this case compared to the previous cases (see Fig.\ref{fig_energies}).   

A similar test of introducing the magnetic structures of CoToB in an isentropic environment was performed. We find the same kind of results as in sect. \ref{sect_cstsbisen}. The interaction between the loops while they rise is again quite similar to the convective case, the emerging region is again very circular and the necklaces very prominent at the edges of the large upflow created by the rising magnetic field.

\subsection{Radial field and fluxes in the different cases}
\label{sect_flux}

To get more quantitative estimates of the typical field strength and flux contained in the emerging regions and compare with observed sunspots, we choose to focus on the most realistic cases where the initial toroidal magnetic field was equal to $6\times10^4 \rm G$. These cases are denoted CsTsBw, CoTsBw, CsToBw and CoToBw. In this situation, the convective motions play a significant role on the rising structures (see Sect.\ref{sect_csts}) and the typical value of the emerging radial field is similar to what is observed in solar active regions, i.e. of the order of a few kiloGauss. \lau{We note however that the top of our computational domain is still about $28 \rm Mm$ below the photosphere and that  large variations of field strength and fluxes could occur within these last $4\%$ of the solar radius. Consequently, we focus here more on the time evolution of these quantities than on their absolute values.}

\begin{figure*}[h!]
	\centering
	\includegraphics[width=8.5cm]{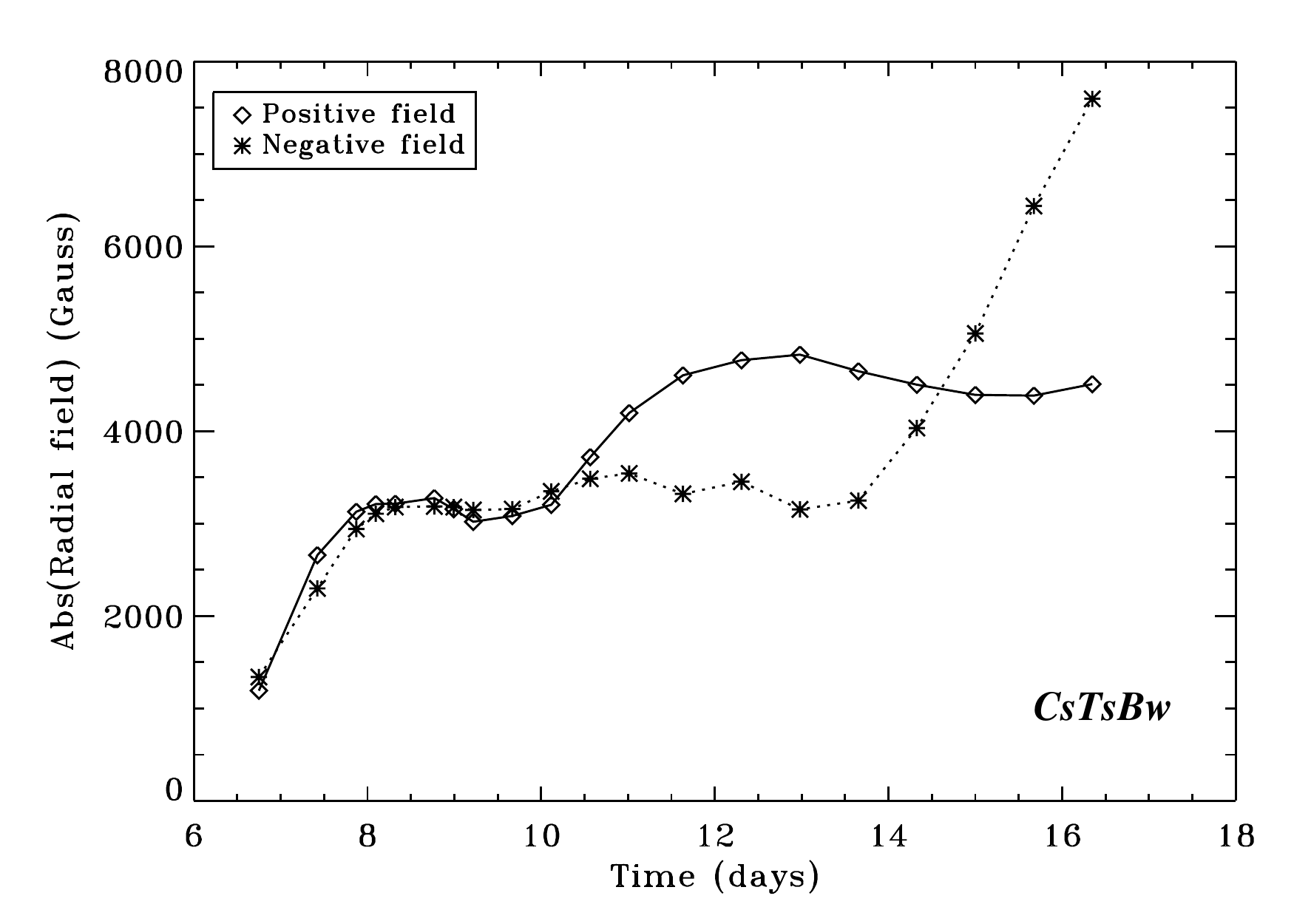}
	\includegraphics[width=8.5cm]{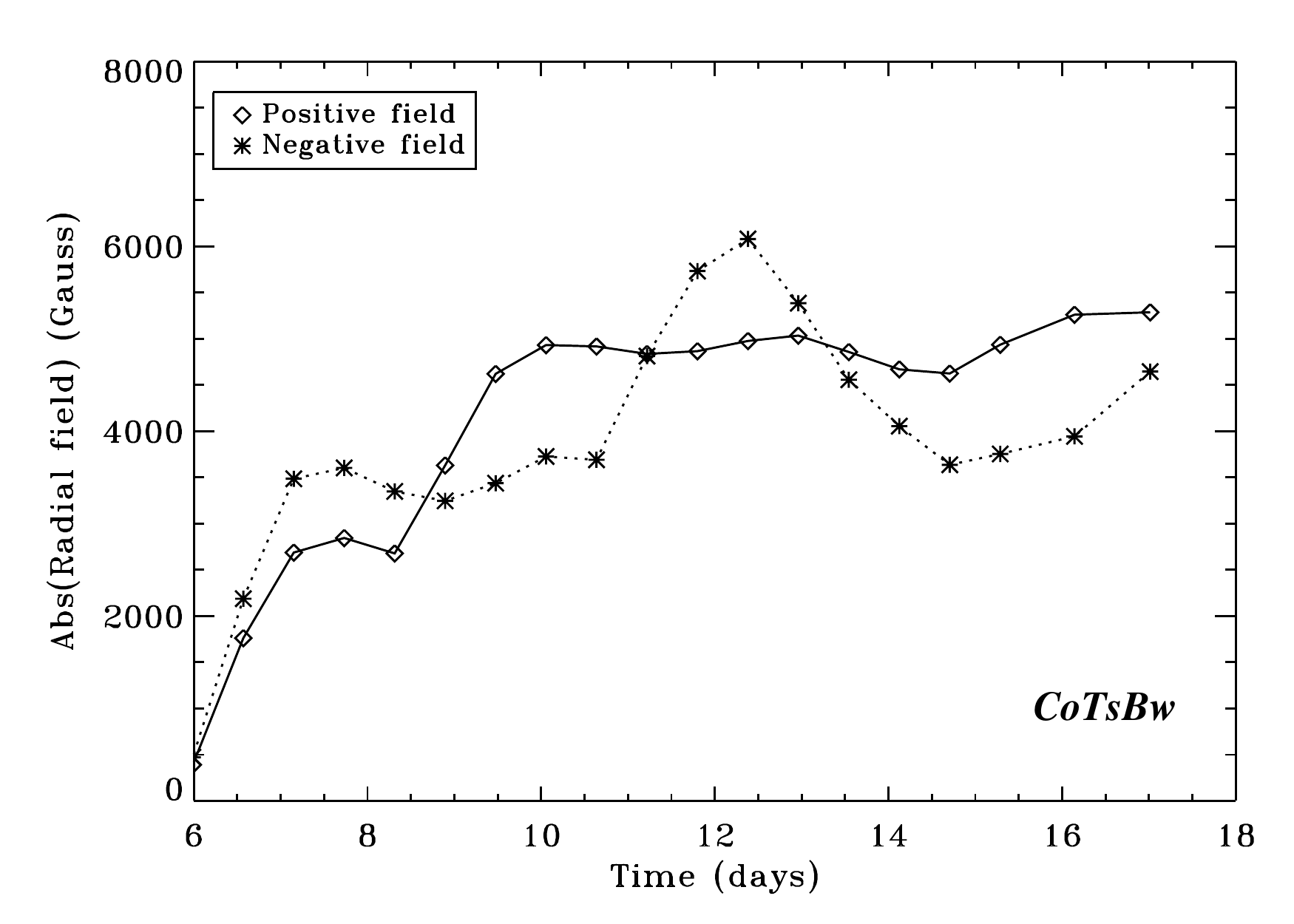}
	\includegraphics[width=8.5cm]{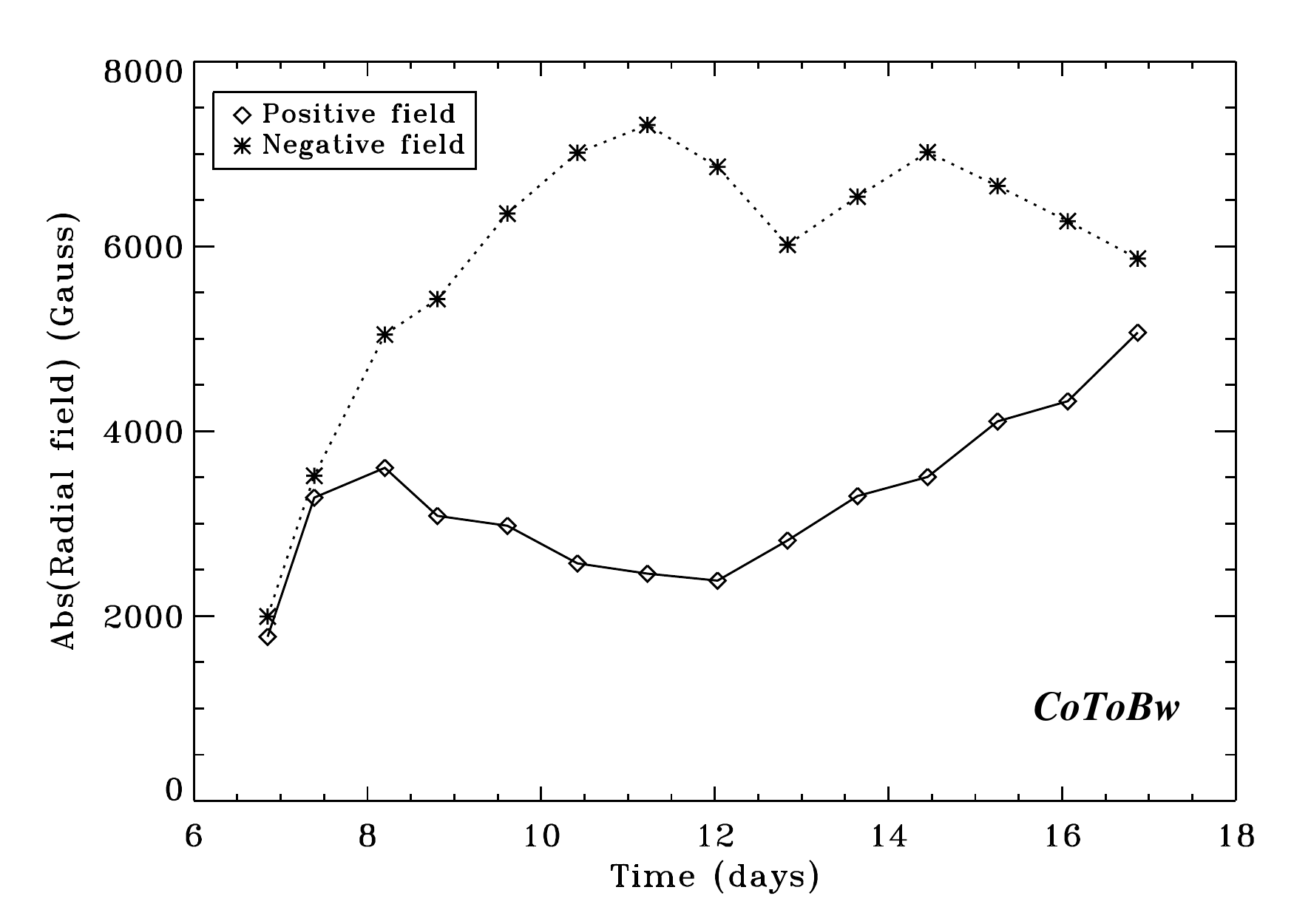}
	\includegraphics[width=8.5cm]{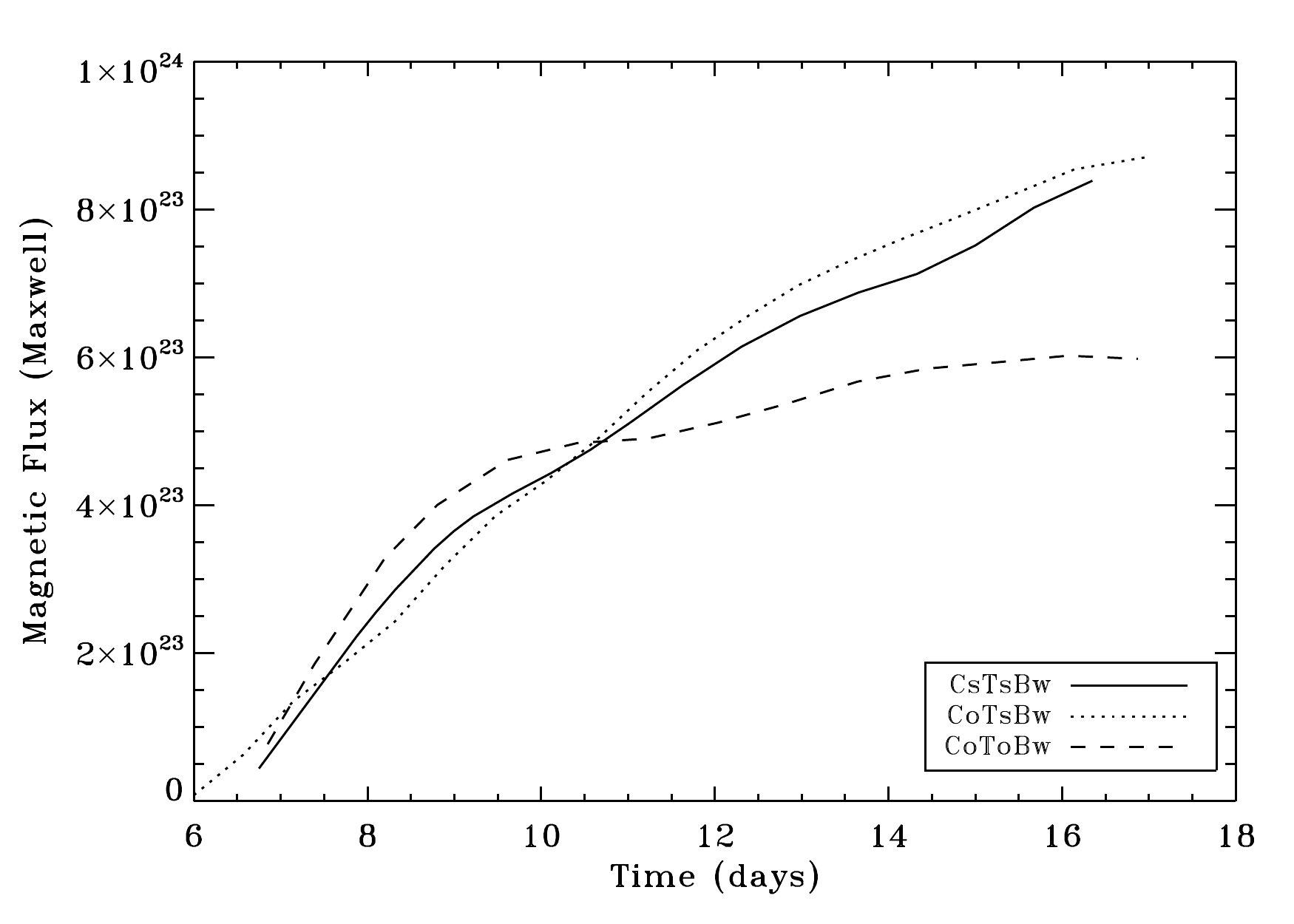}
	\caption{Maximum of the absolute value of the radial magnetic field in both polarities in the emerging region, from the time of emergence to about 17 days, for CsBsTw (top left), CoTsBw (top right) and CoToBw (bottom left). Bottom right panel: flux in the emerging region, from the time of emergence to about 17 days, for CsTsBw (plain), CoTsBw (dotted) and CoToBw (dashed).}
	\label{fig_br}
\end{figure*}

Figure \ref{fig_br} shows the temporal evolution of the value of the maximum radial field contained in the positive and negative polarities of the emerging bipolar regions at $r=0.93 R_\odot$, for CsTsBw (top left panel), CoTsBw (top right) and CoToBw (bottom left). The first two panels show a rather similar evolution of the radial magnetic field, with a sharp increase during the first days of emergence of the intensity in both polarities and then a plateau at around $B_r \approx  4-5 \rm kG$. The time evolution is then flatter but when the magnetic field starts to be significantly advected towards the down flow lanes at the periphery of the large up-flowing convective cells, compression of the field lines can be observed through the sudden increase of the field intensity in one of the polarities. This is visible in particular from day 14 on the top left panel for the negative polarity and on the top right panel for the negative polarity at around $t=11$ days. The increase of these negative $B_r$ concentrations are related to the squeezing of the field lines in the narrow lane at the periphery of the emerging regions. The evolution is however quite different for CoToBw where full reconnection occurs and where we found that the two main concentrations of $B_r$ are not organized as well-defined spots, contrary to the previous cases. The intensity of the radial magnetic field in the two polarities is then dictated mostly by the structure of the convective cells they are embedded in. In particular, the negative polarity is quickly advected towards the peripheral down flow lanes where it is strongly compressed and where a local increase of the magnetic intensity is thus produced. The positive polarity is more diffused, the compression effect is thus reduced and the maximum radial field thus levels off at a value roughly twice as less as the maximum in the negative polarity. This is consistent with the structure of $B_r$ which was observed at $t=10$ days in the last panel of Fig. \ref{fig_shslfull} where the compression of the negative polarity is much more pronounced than that of the positive one.

%\begin{figure*}[h!]
%	\centering
%	\includegraphics[width=8cm]{fig21.pdf}
%	\caption{Flux in the emerging region, from the time of emergence to about 17 days, for CsTsBw (plain), CoTsBw (dotted) and CoToBw (dashed).}
%	\label{fig_flux}
%\end{figure*}

The temporal evolution of the positive flux in the emerging active regions is shown on the bottom right panel of Fig.\ref{fig_br}. Again, CsTsBw and CoTsBw have a very similar behavior, with roughly the same amount of flux coming from the emerging region at $r=0.93 R_\odot$ and the same growth rate. If we calculate a rough estimate of the initial flux contained in our flux tubes of radius $a$ and field strength $B_0$, we find values of the order of $7.5\times 10^{23} \rm Mx$ for each loop, \lau{corresponding to a total flux of $1.5\times 10^{24} \rm Mx$}. At $r=0.93 R_\odot$, the flux in the emerging region approximately reaches half this value in CsTsBw and CoTsBw. This implies that half of the flux remains buried in the convection zone, which is consistent with the fact that only a portion of the loop was made buoyant initially so we do not expect the whole magnetic flux to be contained in the emerging region. CoToBw is again peculiar since the initial increase of the flux is faster than in the two previous cases but saturates at a value of around $6\times 10^{23} \rm Mx$, well below the flux initially contained in our tubes. Because of the additional reconnections which occurred between the loops in this case compared to CsTsBw and CoTsBw, part of the flux is lost during rise. This is not in contradiction with the results presented in Fig.\ref{fig_br} where the maximum radial field was in fact stronger than in the two other cases. The flux is reduced because of the much more elongated and fine-scaled structures created in this case, as seen in Fig.\ref{fig_shslfull}. \lau{We note again that these values are about 10 times higher than the flux in the strongest solar active regions \citep{Aulanier13} and that reducing the initial loop radii only by a factor 4 could possibly produce realistic simulations of these very strong active regions. Again, directly comparing with observations was not the purpose of this work since the very top of the convection zone is not included in our domain.}

\section{Discussion: creation of complex active regions}

One of the main motivations of this work is to identify the possible origin of complex active regions. The definition of ``complexity" of an active region is however rather unclear.
We here choose to focus on the radial magnetic field and its relations to the radial current. Of particular interest are firstly the possible emergence of non-bipolar active regions and secondly the degree of neutralization of the electric currents in each polarity of the emerging region. It has been shown that the non-neutralized currents in active regions are at the origin of eruptive events at the solar surface \lau{\citep{Forbes10, 2015ApJ...810...17D,Kontogiannis17}}. It is however still an open question to understand where the non-neutralization comes from (see for example  \cite{Torok14} and \cite{2015ApJ...810...17D} for further discussions on the subject). It has been argued that this could originate from both the process of flux emergence and of horizontal flows twisting and shearing the magnetic structures. In our simulations, these mechanisms are all present so that non-neutralized currents should exist. Moreover, we have the possibility here to study if the various interactions between loops will produce a different quantity of non-neutralized currents.

\subsection{Multipolar active regions}

In our various cases, we obtain different configurations for the emerged radial magnetic field. To connect with observations, we can try to identify cases for which the active region does not possess a well-defined bipolar structure but on the contrary exhibits some multipolar characteristics. Going back to Figs \ref{fig_shslbounce}, \ref{fig_shslmerge}, \ref{fig_shslmergew}, \ref{fig_shslmergeisen} and \ref{fig_shslfull}, we can isolate some cases with a multipolar radial magnetic field. The upper panels of each of these figures allow us to see that this complexity evolves during emergence. For example in CsTsBw (Fig.\ref{fig_shslmergew}), the loops merge quite late during their rise through the convection zone and thus at time $t=7$ days, the 2 separate structures can still be seen on the radial field map, contrary to CsTsB (Fig.\ref{fig_shslmerge}) in which the loops merged at the very first stages of their evolution. As a consequence, the structures will first appear multipolar and while the emergence and the merging occur, this complexity tends to vanish so that a clear bipolar structure is observed at the later time $t=10$ days. Even if some opposite polarity regions still exist inside each spot, this region is thus quite simple in terms of $B_r$. However, we show here that a region going from multipolar to bipolar during its emergence could indicate that a merging between 2 separate structures took place in the solar interior, which could be of interest for observational studies in which the early time evolution of ARs is followed. The most complex structure obtained here is in CoTsBw (Fig.\ref{fig_shslbounce}). Of course in this case, 2 separate regions emerge, naturally creating a multipolar structure if some overlap exists between the field concentrations. If we ignore the region at higher latitude, it is still interesting to see that the other structure exhibits a pattern more complex than what is obtained in the merging cases. This could be due to the fact that the merging in CsTsBw allows a re-concentration of magnetic field that makes it less sensitive to the surrounding convective motions. The advection by convective flows of weak fields will thus also affect the multipolarity of the radial field. In CoTsBw, we note that the structure was already multipolar when it started to emerge (at $t=8.5$ days) and could thus be distinguished from the merging case CsTsBw. 

The bottom right panel of Figs \ref{fig_shslbounce}, \ref{fig_shslmerge}, \ref{fig_shslmergew}, \ref{fig_shslmergeisen} and \ref{fig_shslfull} show the distribution of the radial current $J_r$ at time $t=10$ days across the emerged magnetic structure. Superimposed are contours of the radial magnetic field at $2\%$ and $80\%$ of its minimum and maximum values. The weak contours enable us to identify the extent of the whole positive and negative polarities. At the boundary between those contours, we thus observe the polarity inversion line (PIL) which also gives some information about the complexity of the region. The strong contours locate the main field concentrations of each sign. Again, it is clear that the case where the PIL is sheared the most is CoTsBw (for the low latitude region). In CsTsB and CsTsBw, the merging cases, tracing the PIL allows us to exhibit a region of negative polarity in the dominantly-positive polarity at around $50^o$ latitude and $95^o$ longitude. This opposite polarity is the result of the merging between the two loops and naturally produces a complexity in the AR. To get another insight in the complexity of the emerged field, let us now focus on the radial current.

\subsection{Non-neutralized currents}

As seen again in the bottom right panel of Figs \ref{fig_shslbounce}, \ref{fig_shslmerge}, \ref{fig_shslmergew}, \ref{fig_shslmergeisen} and \ref{fig_shslfull}, in all cases the distribution of $J_r$ is rather complex, with both signs existing in each polarity. Generally and as expected, the strongest currents are located at the periphery of the emerged region, where the strongest gradients of magnetic field lie. This peripheral current is usually of opposite sign of the one located in the bulk of the field concentrations. The latter is reminiscent of the so-called ``direct" currents located in the center of individual idealized flux tubes while the peripheral currents are similar to the ``return" currents surrounding the direct ones. In cases CsTsB and CsTsBw (figures \ref{fig_shslmerge} and \ref{fig_shslmergew} where a merging of the loops has occurred), the domains occupied by the positive and negative polarities have a complex geometry, with islands of opposite signs present in both polarities. In those merging cases, we anticipate that some regions of positive and negative currents will dominate over individual polarities. As stated in the introduction, the subsequent ``non-neutralized" currents in each polarity will give us some insight on the eruptive skills of our emerging magnetic structures.

To have a quantitative measurement of the ``non-neutralized" currents in each polarity, we calculate the following quantities:

$$
J_{+}=\frac{\iint_{pp} J_r \,\, ds}{\iint_{pp} \vert J_r \vert \,\, ds}
$$
$$
J_{-}=\frac{\iint_{np} J_r \,\, ds}{\iint_{np} \vert J_r \vert \,\, ds}
$$

where the surface integration is performed on the positive polarity ($pp$, i.e. wherever $B_r$ is positive) for $J_{+}$ and on the negative polarity ($np$, i.e. wherever $B_r$ is negative) for $J_{-}$. These quantities will give us the percentage of radial current which is not cancelled by the other sign in both polarities. These values will also have a sign, indicating which sign of $J_r$ is dominant in each polarity. 

\begin{deluxetable*}{c|ccc|ccc}

\tablehead{
\colhead{} & &\multicolumn{2}{c}{t=8.5 days} & &\multicolumn{2}{c}{t=10 days} \\
\hline
\colhead{Cases}& &\colhead{$J_{+}$} & \colhead{$J_{-}$} & &\colhead{$J_{+}$} & \colhead{$J_{-}$}
}

\startdata
CsTsB&  & 39.2 & -39.2 & &26.5 & -25.6   \\
CsTsBw& & 52.7 & -64.8 & &40.6 & -48.5   \\
\hline
CoTsBw& &12.9 & -7.0 & &-11.0& 12.4   \\
CoToBw& &11.2 & -11.1 & &9 & -6.9   \\
\enddata

\caption{Percentage of non-neutralized radial currents in each polarity in the 4 convective cases, at 2 different times and at $r=0.92 R_\odot$. Data are given in percents. The sign indicates which sign of the non-neutralized radial current dominates.}
\label{tab:table}
\end{deluxetable*}

The values of $J_{+}$ and $J_{-}$ are indicated in Table \ref{tab:table} for CsTsB, CsTsBw, CoTsBw and CoToBw at 2 different times $t=8.5$ and $t=10$ days. If we focus on time $t=8.5$ days, the most striking feature is that the cases without merging all possess less than $13\%$ of ``non-neutralized" currents while the merging cases exhibit values above $39\%$, and as high as almost $65\%$ for the case where the initial field was weaker. This was indeed anticipated when looking at the snapshots of the radial currents in Figs \ref{fig_shslbounce}, \ref{fig_shslmerge}, \ref{fig_shslmergew}, \ref{fig_shslmergeisen} and \ref{fig_shslfull}. These differences are still present at the later time $t=10$ days but in a milder proportion, even though CsTsBw still exhibits large quantities of ``non-neutralized currents" at that time, almost $50\%$ in each polarity. The global decrease of the values in time for all cases is interpreted as an effect of magnetic diffusion which become important when the flux tubes reach the top of the computational domain where convective motions significantly shuffle the field lines.

We can now focus on the behavior of each case. First, in CoToBw, reconnections occur between peripheral and axial field lines and these efficient reconnections seem to wipe out most of the current sheets which form during the loop merging. Then we find that CoTsBw possesses a small degree of non-neutralization. In this case, the loops only bounce against each other and the structure of the currents in each loop should thus remain quite similar to the case of two individual loops evolving independently of each other. As seen in Fig.\ref{fig_lotw}, a strong current sheet at the boundary between the two loops appears for CsTsB but not for CoTsBw.  This sheet will be the location of reconnection and thus strong annihilation of return currents, that explains the existence of non-neutralized currents in CsTsB. When this sheet does not occur (like in CoTsB), the degree of non-neutralized currents is then expected to be smaller. We note that CsTsBw, with a weaker field, has a higher degree of non-neutralization. This is a case where convective motions and thus shearing and twisting could have stronger effects. The ability of these convective motions to generate a larger amount of non-neutralized currents is interesting and should be considered in local models where a detailed understanding of the origin of non-neutralized currents is more tractable. We thus find here that the merging process between the 2 loops is favorable for the creation of currents of one particular sign in each polarity, contrary to what was found in CoToBw where the reconnections could have been too efficient and thus result in opposite-sign currents quickly annihilated. The fact that the merging cases, and especially CsTsBw, where the effects of convective motions are strongest, possess a large proportion of ``non-neutralized" currents is another evidence that these cases produce a complex emerging magnetic structure, with a \lau{potentially} high likelihood for eruptions. \lau{To investigate the effects of magnetic diffusion in CsTsBw, we performed the same simulation but with a magnetic Prandtl number $P_m=4$, i.e. with a magnetic diffusivity reduced by a factor 4. We find some additional small-scale emerging structures which were diffused away in the $P_m=1$ case and overall stronger values for the radial magnetic field and current are obtained. However, the global reconnection still occurs and the amount of non-neutralized currents is still high in this case: at time $t=8.5$ days, $J_{+}$ and $J_{-}$ are respectively equal to $43$ and $-54$ (to be compared with $52.7$ and $-64.8$ for the $P_m=1$ case) and at $t=10$ days, they reach the values $35$ and $-40$ (to be compared with $40.6$ and $-48.5$ for the $P_m=1$ case). These cases, where large amounts of non-neutralized currents exist,} are thus interesting candidates to follow in time, in particular their emergence in an overlying atmosphere where further reconnection and relaxation could occur.

\section{Conclusion}

In this work, interactions between magnetically buoyant loops embedded in a convective shell were investigated, with a particular focus on the morphology of subsequent emerging regions. Different cases were investigated, mimicking the different situations which could occur at the base of the convection zone where these buoyant structures are assumed to be formed. In agreement with previous studies in Cartesian geometry and without convective motions, we find that the interactions between the loops are dictated mostly by the possible reconnections between field lines. \lau{When the loops are initiated with antiparallel field lines at their periphery and when these are sufficiently close, they can merge. On the contrary, they bounce against each other when the field lines are initially parallel to each other.} An extreme case is one where both the peripheral field lines and the axial field lines of one loop are anti-parallel to the ones of the other loop. In this case, efficient reconnections take place, producing a much more dispersed radial field at the surface instead of well-defined bipolar structure.

The level of complexity obtained in the emerging magnetic field is also quite different between those various cases. To quantify this level of complexity, we chose first to investigate the maps of emerging radial fields. Multipolar structures and sheared polarity inversion lines can be found in the merging cases as well as in the bouncing cases where convection has been sufficiently active. We find that the apparent complexity decreases in the merging cases when the interaction between the loops occur close to the surface of our domain. We then investigated the formation of radial currents close to the top of our domain and focused on the degree of non-neutralized currents appearing in each polarity. This quantity is indeed thought to be crucial to characterize the ability of magnetic field configurations to give rise to eruptive events like flares or CMEs. We find that the merging cases are very efficient at producing a large level of non-neutralized currents, especially when convective motions significantly act on the emerging structure. Some additional calculations are needed to understand the exact origin of non-neutralized currents in these particular cases. However, this already confirms that twisting and shearing motions, as well as strong interactions between emerging structures could indeed produce complex magnetic fields which will then tend to relax to simpler configurations by releasing large amounts of their free energy.

Our study extends the previous studies of interacting idealized flux tubes by incorporating the effects of convection, large-scale flows and spherical geometry. We moreover get to the conclusion that individual buoyant structures merging during their rise to the solar surface are likely to produce complex magnetic configurations favorable for eruptive events. This would be of strong interest to study the further evolution of such emerging fields in a realistic atmosphere where the triggering of eruptions and the magnetic field relaxation could also be investigated. The definition of complexity is however difficult to determine precisely. We decided here to focus on the non-neutralized currents in each polarity. Another quantity of interest to measure complexity could also be $\alpha={\bf J \cdot B}/B^2$, which can be related to the observed $J_z/B_z$ where $\bf e_z$ is the direction of the line of sight. \lau{This quantity has been used to analyse observations of photospheric magnetic fields for example in \cite{Pevtsov95} and later in the statistical studies of \citet{Leka03a, Leka03b}}. The monitoring of $\alpha$ in our calculations was performed, without producing very clear differences between our various cases but we plan to discuss those questions in further studies. For now, our present study could enable us to determine from observed maps of line-of-sight currents, the possible magnetic structure below the surface or which type of interaction could have happened before emergence.

Finally, we saw that reconnection may be favorable to building complex magnetic configurations. It would thus be necessary to study the evolution of interacting buoyant loops in a magnetized environment where reconnections could also occur with the surrounding smaller-scale magnetic field. This will also be the subject of future work.

\acknowledgments

We wish to thank M. Linton for very useful discussions during the Flux Emergence Workshop 2017 in Budapest, the main organizers of which (T. T{\"o}r{\"o}k and K. Petrovay) we also want to thank. L.J. wishes to acknowledge the support from the Institut Universitaire de France (IUF). \lau{A.S.B. acknowledges funding by CNES through a Solar Orbiter grant, by INSU/PNST and by the ERC through the Solar Predict grant.}

\bibliography{ms_jouve}

\end{document}